\pdfoutput=1
\documentclass[11pt]{article}

\usepackage[final]{acl}

\usepackage{times}
\usepackage{latexsym}
\usepackage{hyperref, enumitem}

\usepackage[T1]{fontenc}
\usepackage[utf8]{inputenc}

\usepackage{microtype}

\usepackage{inconsolata}

\usepackage{graphicx}

\usepackage{multirow}
\usepackage{booktabs} 
\usepackage{amsmath}
\usepackage{bm}
\usepackage{makecell}
\usepackage{xspace}
\usepackage{xurl} 

\newcommand{\systemName}{\texttt{Action-Design Scoring}\xspace}

\setenumerate[1]{itemsep=0pt,partopsep=0pt,parsep=0pt,topsep=0pt,leftmargin=0pt}
\setitemize[1]{itemsep=0pt,partopsep=0pt,parsep=0pt,topsep=0pt,leftmargin=12pt}
\setdescription{itemsep=0pt,partopsep=0pt,parsep=0pt,topsep=0pt,leftmargin=0pt}

\usepackage{listings}
\lstdefinestyle{promptstyle}{
    basicstyle=\ttfamily\small,        % Typewriter font, small size
    breaklines=true,                   % Enable line breaking
    frame=single,                      % Single line frame around the listing
    framerule=0pt,                     % Remove the frame border
    backgroundcolor=\color{gray!10},   % Light gray background
    captionpos=b,                      % Caption position at the bottom
    numbers=none,                      % Disable line numbers
    xleftmargin=0em,                   % Left margin
    xrightmargin=0em,                  % Right margin
}

\title{Automating eHMI Action Design with LLMs \\ for Automated Vehicle Communication}

\author{
 \textbf{Ding Xia\textsuperscript{1}},
 \textbf{Xinyue Gui\textsuperscript{1}},
 \textbf{Fan Gao\textsuperscript{1}},
 \textbf{Dongyuan Li\textsuperscript{1}}
 \textbf{Mark Colley\textsuperscript{1,2}},
 \textbf{Takeo Igarashi\textsuperscript{1}}
\\
 \textsuperscript{1}The University of Tokyo,
 \textsuperscript{2}University College London
\\
\texttt{\{dingxia1995, fangao0802, lidy94805\}@gmail.com}, \\
\texttt{gui-xinyue@g.ecc.u-tokyo.ac.jp}, 
\texttt{m.colley@ucl.ac.uk}, 
\texttt{takeo@acm.org}
}

\begin{document}
\maketitle

\begin{abstract}
% For the discussion page, refer to the \href{https://docs.google.com/document/d/1NOxqIFv9Gx2r6Se7Kt-0IoQw8Ah2DjDPCt711LVE4CU/}{GUI-XIA document}.

The absence of explicit communication channels between automated vehicles (AVs) and other road users requires the use of external Human-Machine Interfaces (eHMIs) to convey messages effectively in uncertain scenarios.
Currently, most eHMI studies employ predefined text messages and manually designed actions to convey these messages, which limits the real-world deployment of eHMIs, where adaptability in dynamic scenarios is essential.
Given the generalizability and versatility of large language models (LLMs), they could potentially serve as automated action designers for the message-action design task. To validate this idea, we make three contributions: (1) We propose a pipeline that integrates LLMs and 3D renderers, using LLMs as action designers to generate executable actions for controlling eHMIs and rendering action clips. (2) We collect a user-rated \systemName dataset comprising a total of 320 action sequences for eight intended messages and four representative eHMI modalities. The dataset validates that LLMs can translate intended messages into actions close to a human level, particularly for reasoning-enabled LLMs. (3) We introduce two automated raters, Action Reference Score (ARS) and Vision-Language Models (VLMs), to benchmark 18 LLMs, finding that the VLM aligns with human preferences yet varies across eHMI modalities.
% Our results offer potential solutions for current eHMI studies and suggest possible directions for future research.
\footnote{The source code, prompts, Blender scenarios, and rendered clips are available at \url{https://github.com/ApisXia/AutoActionDesign}}

\end{abstract}

\section{Introduction}
\label{sec:intro}

\begin{figure}[ht]
    \centering
    \includegraphics[width=\linewidth]{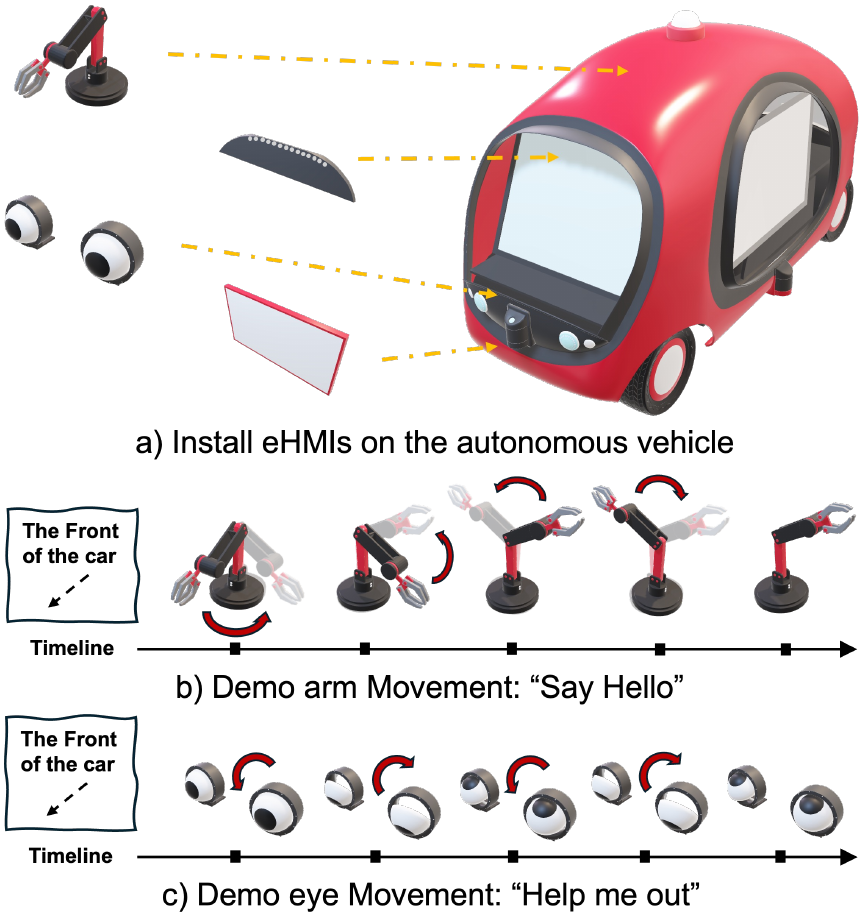}
    \caption{Setup illustration and action demos. a) Four types of eHMIs are installed on the vehicle separately; b) Demo actions of the arm convey the message: ``Say Hello''. The shaded action indicates the subsequent status; c) Demo actions of  the eye: ``Help me out''.}
    \label{fig:demo_cases}
\end{figure}

Automated vehicles (AVs) promise to redefine transportation systems by eliminating human driving errors and optimizing traffic flow~\cite{fagnant2015preparing}. 
However, the absence of a human operator disrupts road interactions, as drivers no longer exchange contextual cues (e.g., eye contact and gestures) to negotiate ambiguous scenarios~\cite{10.1145/3706598.3714187}. To bridge this gap, external Human-Machine Interfaces (eHMIs) have emerged as mediators, conveying AV intent (e.g., yielding, turning) to other road users, such as pedestrians, cyclists, and human drivers~\cite{dey2020taming, colley2020design}.
These interfaces use diverse forms, such as displays~\cite{al2024light, lim2022ui}, LED strips~\cite{dey2020color}, projections~\cite{eisma2019external}, and attached robots~\cite{gui2024shrinkable}, to convey vehicle intentions by text, signals, or non-verbal motions.

Current eHMIs are usually designed and analyzed with predefined text messages (e.g., ``Please stop.'', ``I am worried.'') with scenario information (e.g., ``pedestrian is crossing the road'', ``robot is stuck in snow'') and manually designed actions to perform these messages, as shown in \autoref{fig:demo_cases}.   
This restricts the real-world deployment of eHMIs because dynamic interactions demand adaptable communication strategies~\cite{dey2020taming}. Therefore, developers must design actions for all possible messages that AVs might need to communicate to other road users.
This process is time-intensive, costly, and significantly limits the scalability of eHMIs in practical scenarios~\cite{vehicles6030061}.

Recently, Large Language Models (LLMs) demonstrate generalizability and versatility in multiple tasks, such as reading and answering questions~\cite{radford2019language}, as well as pattern following~\cite{mirchandani2023large}, suggesting that they may serve as suitable automated action designers for eHMIs. However, it is unclear whether the application of LLMs for eHMIs is feasible and useful, leading to our research question (RQ): 
\begin{quote}
\textit{To what extent do pretrained LLMs achieve parity with human designers in designing eHMI actions that are understandable to road users?}
\end{quote}
Answering our RQ involves three key challenges. \textbf{First}, there is no systematic pipeline for translating specified messages into executable action sequences for eHMIs. \textbf{Second}, there is a lack of high-quality datasets for testing and improving the translation of eHMI messages into action sequences. \textbf{Third}, there is no commonly used benchmark to compare different methods for designing and evaluating eHMI actions fairly.

Therefore, \textbf{first}, we propose a pipeline that integrates LLMs and 3D renderers. To adapt LLMs for controlling eHMIs, we draw inspiration from LLM-based robot action planning~\cite{garrett2021integrated, zitkovich2023rt}, which utilizes LLMs as action designers to generate a series of executable actions to actuate robotic motors. 
\textbf{Second}, we introduce a user-rated \systemName dataset. The dataset comprised eight intended messages for the eHMI to convey by analyzing traffic scenarios, and selected four representative eHMI modalities frequently discussed in eHMI research. We collected messages from previous eHMI studies~\cite{chang2022can, gui2022going, gui2024scenarios} and designed new ones based on message types~\cite{colley2020design} to enrich the variety. For each message-modality pair, we generated ten actions: eight produced by LLMs (GPT-4o~\cite{achiam2023gpt}, Sonnet 3.5~\cite{Anthropic2024}, Gemini 2 Flash~\cite{DeepMind2024}, and GPT-o1~\cite{openaiO1preview}) and two designed by human experts. These actions were rendered using Blender version 4.3~\cite{blenderorg}, resulting in 320 video clips. We conducted a video-based user study with human participants. They evaluated the understandability of the LLM-designed actions by measuring the consistency between the intended messages and perceived meanings. The \systemName dataset provides averaged human scores for each action, enabling a comparative benchmark for existing LLMs. 
\textbf{Third}, we introduce the Action Reference Score (ARS), which uses the similarity between the newly designed actions and those rated in our dataset. Additionally, we discussed the potential of Vision-Language Models (VLMs) to serve as human-like raters. Then, we benchmark 18 LLMs on this task.

\textit{Contribution Statement:}
This work proposes the first complete pipeline, along with a comprehensive dataset and benchmark for evaluating eHMIs. Beyond these core contributions, we also share several noteworthy insights as follows: 
\begin{itemize} 
    \item Pretrained LLMs can achieve a close human-level action design capability (see \autoref{sec:result_dataset_analysis}).
    % \item VLM rater aligns with human scoring preferences, but is influenced by eHMI modalities.
    \item VLM rater matches human preferences but varies across eHMI modalities (see \autoref{sec:result_vlm_evaluation}).
    \item Reasoning-enabled LLMs demonstrate better performance in our task (see \autoref{sec:benchmark}).
\end{itemize}

\section{Related Work}
\label{sec:relate}

% This section introduces rule-based eHMI and LLM-based robot action planning.

\subsection{Rule-Based eHMI Action Planning}

Currently, eHMI action planning generally follows a fixed design approach, in which human designers establish behavioral rules based on the specific features of different eHMI modalities. For example, in text- and icon-based eHMIs, designers create content referencing traffic regulation signs or standard messages~\cite{eisele2022effects,eisma2021external}. 
In color- and light-band-based eHMIs, they design the content relying on human intuitive empathy and empirical evaluation with colors and blinking frequencies~\cite{bazilinskyy2019survey, dey2020color}. For anthropomorphic eHMIs, such as eyes or arms, designers mimic nonverbal communication cues drawn from common human-human interactions~\cite{mahadevan2018communicating,ochiai2011homunculus}.
Most recently, \cite{10.1145/3706598.3714187} proposes using Human-In-The-Loop Multi-Objective Bayesian Optimization to create appropriate eHMIs. 
However, these eHMI action design only works as part of a case study to validate new eHMI modalities, which does not encourage the emergence of action design methods before our paper.

In summary, traditionally, experts have observed real-world examples to derive design rules for guiding eHMI action planning. However, different eHMI modalities vary in expression. Low-expressiveness eHMIs, such as arrow icons, are relatively simple because they convey static directional cues, making it easier to define behavioral rules~\cite{fridman2017walk}. Highly expressive eHMIs can produce complex actions and communicate richer messages~\cite{chang2024must}. However, determining behavioral rules for such modalities is challenging due to the increased intricacy and variability of their expressions~\cite {gui2023field,de2022external}. 
Unlike previous works, we address this by evaluating LLMs to support eHMI action planning, enabling more complex and dynamic communication.

\subsection{LLMs-Based Robot Action Planning}
Recent LLMs encode vast world knowledge and exhibit the emerging capability for robot action planning~\cite{xiang2024language}. Regarding how LLMs generate actions to actuate robots, existing approaches fall into two main trends: Task and Motion Planning (TAMP)~\cite{garrett2021integrated} and Visual Language Action (VLA) models~\cite{zitkovich2023rt}. 
TAMP methods break down complex instructions into predefined low-level actions (e.g., grasping, moving) to control robots~\cite{chen2024autotamp}. However, for our task, it is difficult to predefine these action categories. We believe that forcing LLMs to choose from rigid action modes limits their ability to design flexible or adaptive actions creatively~\cite{hao2025learn}. In contrast, VLA models fuse robot control actions directly into VLM backbones, providing specific action commands to control each robotic motor~\cite{zitkovich2023rt,kim2406openvla}. However, applying existing VLA models to out-of-scope tasks with different robot settings often requires a large amount of data for finetuning~\cite{qu2025spatialvla}. This contradicts our objective of reducing the labor required by human experts in designing eHMI actions. In this task, we utilize the generalizability and versatility of pretrained LLMs by providing detailed prompts on how to control each modality of eHMI.

\begin{figure*}
    \centering
    % \hspace*{-0.03\linewidth}
    \includegraphics[width=\linewidth]{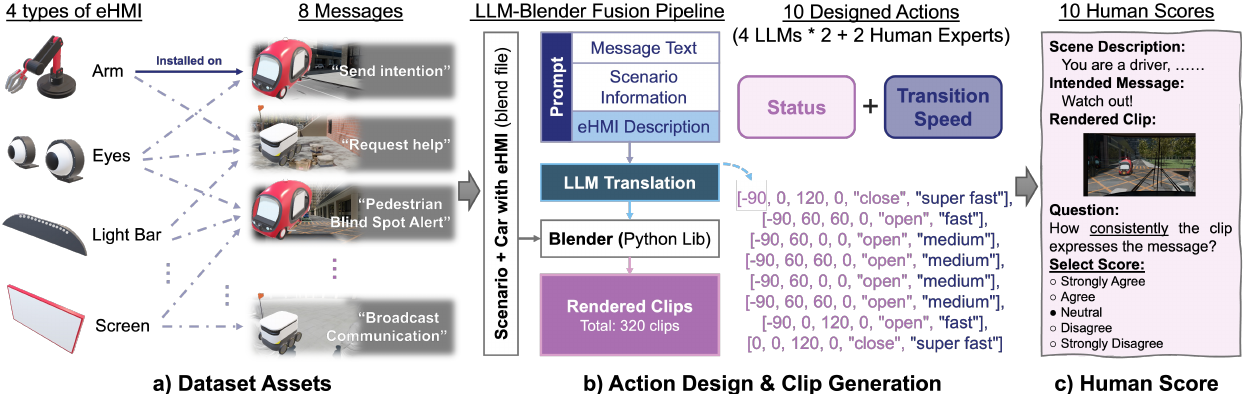}
    \caption{Dataset Asset, Pipeline, and Human Scoring. Dataset assets contain four representative eHMIs and eight intended messages from different interaction types. In the pipeline, we develop eight corresponding Blender scenarios and render actions designed by LLMs or human experts to clips. During the human scoring phase, ten participants evaluate each action clip using a five-point Likert scale.}
    \label{fig:dataset_explain}
\end{figure*}

\section{Methodology}
\label{sec:method}
This section outlines our responses to three key challenges: i) the LLM-Blender Fusion pipeline, ii) the \systemName dataset, and iii) the automated evaluation system for benchmarks. 
These designs serve as a proof of concept for our RQ and offer a systematic approach to developing and evaluating newer LLMs or eHMIs modalities.

\subsection{LLM-Blender Fusion Pipeline}
\label{method:pipeline}
The design of the LLM-Blender Fusion Pipeline (see \autoref{fig:dataset_explain}(b)) unfolds in two steps: i) Designing eHMI actions using LLMs with the provided message text, scenario information, and eHMI description (see details in \autoref{method:ehmi} and \autoref{method:message}), and ii) Rendering the designed actions into corresponding virtual scenarios as video clips in Blender (see \autoref{method:score} for more details).

\subsection{Action-Design Scoring Dataset}

\subsubsection{eHMI Modality Definitions}
\label{method:ehmi}

As shown in \autoref{fig:dataset_explain}(a), four representative eHMIs are selected for analysis, categorized into two types: anthropomorphic (human-like) and non-anthropomorphic~\cite{bazilinskyy2019survey, dey2020taming}. The selection prioritizes dynamic interfaces that use sequential visual cues (e.g., changing brightness, animations) to communicate intent clearly to other road users~\cite{wilbrink2021impact}.
We craft prompts (see Appendix~\ref{appendix:ehmi_describe}) for each eHMI modality, offering detailed guidance to LLMs on what they can control and how to control. Each step of the designed action sequence includes a subsequent status and a transition speed, such as [angle1, angle2, ..., ``fast''].

The descriptions for each status of different eHMI modalities are shown as follows:
\begin{description}
    \item[Eyes.] Robotic eyes are mounted on the front of the autonomous vehicle. The pupil's position is specified using polar coordinates: the angle spans $[0^{\circ}, 360^{\circ}]$ (starting from ``up'' and moving counterclockwise), and the distance spans $[0, 1]$, where $0$ denotes the center and $1$ is the edge~\cite{chang2022can,gui2022going}.
    \item[Arm.] A robotic arm is mounted on the top of the vehicle. It is composed of five components, each of which is connected by single-axis rotational joints. The five movable components (shoulder, upper arm, forearm, hand, and fingers) are required to operate within limited ranges~\cite{gui2024shrinkable}.
    \item[Light Bar.] A light bar contains 15 lights arranged in an arc fixed on the front top of the autonomous vehicle. Each light can be either ``on'' or ``off'', with uniform brightness and color~\cite{dey2020color}.
    \item[Facial Expression.] A screen located at the front of the vehicle displays a sequence of facial expressions to convey messages. The available facial expressions are selected from a set of emojis~\cite{al2024light,dey2020taming}.
\end{description}

Regarding transition speed, we offer three options (e.g., ``slow'', ``medium'', and ``fast''). Additionally, we include a ``super fast'' option to quickly reset the eHMI to its initial status, ensuring continuity when switching between different meanings in an action sequence. In our practical experiments, we find that providing the concept of transition speed, rather than stating specific times like ``1 second,'' gives LLMs a more accurate sense of timing for designing actions. This approach is beneficial because LLMs may not inherently understand the physical scale of the eHMI (e.g., its size or mounting height) or its spatial relationship to other road users, which could lead to ambiguity in interpreting the real-world impact of transition speeds.

\subsubsection{Message Set Design}
\label{method:message}

The communication relationships can be categorized into four types: one-to-one, one-to-many, many-to-one, and many-to-many, where the former (e.g., AVs equipped with an eHMI) interacts with the latter (e.g., pedestrians, cyclists)~\cite{colley2020design}. However, evaluating the collaboration of multiple AVs (e.g., many-to-one, many-to-many) falls outside the scope of this work. Instead, we focus on one-to-one and one-to-many relationships. For one-to-one interactions, we further distinguish between first-person perspectives, where the communicator transmits messages about the AV's state or intent (e.g., ``Help me out!''), and third-person perspectives, where the AV relays information about other road users or environmental conditions (e.g., ``Pedestrian ahead''), based on different perspective taking~\cite{bazilinskyy2019survey}.
% (see \autoref{fig:message_types}).
We collect six messages from previous eHMI studies~\cite{chang2022can, gui2022going, gui2024scenarios} and design two new ones based on message types~\cite{colley2020design} to enrich the variety (see \autoref{tab:ehmi_messages}). Each message includes:
\begin{itemize}[noitemsep]
    \item A message text needs to be conveyed.
    \item Scenario information related to the message. 
    \item A user perspective scenario description for the scoring task. (see Appendix~\ref{appendix:scenario_prompt_avs})
\end{itemize}

\begin{table*}[!t]
  \centering\footnotesize
  \setlength{\tabcolsep}{4pt}        % optional: tighten columns
  \renewcommand{\arraystretch}{1.2}   % increase row height
  \caption{Eight messages collected or designed based on different communication relationships. Each message contains a message text, scenario information, and a user perspective scenario description (see Appendix~\ref{appendix:scenario_prompt_avs}).}
  \label{tab:ehmi_messages}
  \begin{tabular}{@{}p{4cm}|p{3.4cm}|p{8cm}@{}}
    \toprule
    \textbf{Case} & \textbf{Message Text} & \textbf{Scenario Information} \\
    \midrule
    \multicolumn{3}{@{}l}{\itshape One-to-one (First-person) communication relationships} \vspace{0.0em} \\
    \quad \textbf{Send intention}            
      & “I am unable to pick you up here. Please walk forward in my direction to a suitable pickup spot.”  
      & You are an autonomous taxi that receives a ride request and arrives to pick up the passenger (on the right roadside). Upon arrival, you detect the passenger standing in an area where parking is not permitted within a 5 m radius. \\
    \quad \textbf{Status report}           
      & “I am about to start moving. Please watch out.”  
      & You are a stopped autonomous vehicle parked near a park, positioned just before a crosswalk. A student is approaching and is about to cross to the other side of the road. \\
    \quad \textbf{Request help}            
      & “I am stuck. Could you please help me out?”  
      & You are a delivery robot that has been trapped by a pile of boxes. Feeling eager to free yourself and continue delivering the items to your customer on time, you notice a passerby who sees your situation but hesitates to assist. \\
    \quad \textbf{Refuse help}            
      & “Thank you for your kindness. Please not touch me.”  
      & You are an expensive and fragile delivery robot stuck in the snow. You are programmed that only your owner can repair you. Meanwhile, a passerby notices your predicament and hesitates to offer assistance. \\
    \midrule
    \multicolumn{3}{@{}l}{\itshape One-to-one (Third-person) communication relationships} \vspace{0.0em} \\
    \quad \textbf{Pedestrian Blind Spot Alert}
      & “Please watch out for a vehicle approaching from your left blind spot.”  
      & You are an autonomous vehicle parking near an intersection with no traffic lights. A pedestrian on the opposite side is walking toward the intersection, facing you. A building blocks his view of an approaching bus heading toward the intersection from his left (from your right). \\
    \quad \textbf{Driver Blind Spot Warning}  
      & “Please watch out for the pedestrian approaching from your right blind spot.”  
      & You are an autonomous vehicle parked at an intersection without traffic lights. A bus is approaching from the opposite direction. A pedestrian is about to use the crosswalk on the opposite side, coming from your left. However, a building obstructs the bus's view, so it cannot see the pedestrian approaching from its right.\\
    \midrule
    \multicolumn{3}{@{}l}{\itshape One-to-many communication relationships} \vspace{0.0em} \\
    \quad \textbf{Target Identification}      
      & “I am sending the package only to this person.”  
      & You are a delivery robot tasked with delivering a package to a customer in a crowded area. Currently, three individuals are standing to your left, front, and right. Your recipient is directly in front of you and is taller than you. \\
    \quad \textbf{Broadcast Communication}   
      & “I am about to turn right. Kindly make a way to avoid conflict.”  
      & You are a delivery robot carrying a package in a crowded area. You want to navigate through the crowd and turn right without causing disruptions. \\
    \bottomrule
  \end{tabular}
\end{table*}

\subsubsection{Clip Generation and Human Scoring}
\label{method:score}

In the previous section, we obtain a total of 32 modality-message pairs for each eHMI modality and message type (see \autoref{fig:dataset_explain}(a)). For each pair, we ask four LLMs (GPT-4o~\cite{achiam2023gpt}, Sonnet 3.5~\cite{Anthropic2024}, Gemini 2 Flash~\cite{DeepMind2024}, and GPT-o1~\cite{openaiO1preview}) to design two distinct actions. Additionally, two human experts also complete this task. This process results in a total of 320 actions.

However, it is implausible for human participants to rate these actions solely based on text-based commands. They need to observe the actual movements of the eHMIs to judge the effectiveness of the designed actions in conveying messages. Therefore, we incorporate the rendering process into our LLM-Blender fusion pipeline(see \autoref{fig:dataset_explain}(b)). The rendering assets include eHMI models, vehicle models, and scenarios. For eHMI models, the arm is available under a free license~\cite{sinitsyn2021robotarm}, the eyes are part of a proprietary model~\cite{chang2022can}, and the light bar and screen are self-designed. For vehicle models, the AV model is proprietary~\cite{chang2022can}, while the delivery-robot model is available under a free license~\cite{condra2021starship}. For the scenarios, we design the corresponding 3D environments for different messages using Blender version 4.3~\cite{blenderorg}, using a paid add-on called \textit{The City Generator 2.0}~\cite{citygenerator}. We use a GPU-equipped device (NVIDIA GTX 4070 Ti) to render these 320 actions into clips, achieving 24 FPS and 1080p resolution to ensure an optimal viewing experience for participants. The entire rendering process takes approximately 100 hours, with each 10-second clip taking an average of about 20 minutes to complete.

Then, we invite N=40 participants to score the action clips (see \autoref{fig:dataset_explain}(c)). Each participant receives 80 random clips, along with the intended messages and the corresponding user perspective scenario information (see Appendix~\ref{appendix:scenario_prompt_avs}). They then answer the question: \textit{``How consistently do the eHMI actions express the message?''} The participants rate each action clip using a 5-point Likert scale (1=Strongly Disagree to 5=Strongly Agree)~\cite{joshi2015likert}. In contrast to other annotation methods, such as pairwise ranking, the 5-point Likert scale alleviates the participants' load~\cite{rouse2010tradeoffs, mantiuk2012comparison} and reliably reflects their preferences toward different actions~\cite{rankin1980comparison, zerman2018relation}.
In total, we collect 3,200 scores, each action clip rated by ten different participants. We then calculate the average of these scores, resulting in 320 average scores for the clips.

\subsection{Automated Scoring for Benchmarking}
\label{sec:metric}

In the future, one may evaluate the translation capability of messages to actions of novel LLMs. However, employing human participants to score generated actions is expensive and time-consuming. To address this, we propose two substitutes.

\subsubsection{Action Reference Score (ARS)}
\label{sec:ars_metric}
We introduce an Action Reference Score (ARS) that automatically generates a score for a new action by retrieving the most similar actions from our dataset, inspired by existing works for similar purposes~\cite{escudero2023dtwparallel, wilson1997improved}. We use Dynamic Time Warping (DTW)~\cite{muller2007dynamic, salvador2007toward, slaypni2015fastdtw} to compute the similarity between actions. DTW is particularly effective because it calculates similarity even when identical patterns appear at different positions or when sequences vary in length. 
Our approach converts the status of next action steps into numerical values. For example, the angle variable (e.g., $60^\circ$) is transformed into its sine and cosine components to capture its cyclical nature. Similarly, categorical variables (e.g., ``close'') are assigned predefined integer values, and transition times are quantified by assigning ``slow'' as 4, ``medium'' as 3, ``fast'' as 2, and ``super fast'' as 1. In defining the distance function for the DTW algorithm, we assign an equal weight of 1 to numerical, categorical, and temporal elements, normalizing each element's value range to [0, 1].

\subsubsection{Vision-Language Model (VLM) Rater}
\label{method:vlm_rater}
We also evaluate whether the designed actions are contextually appropriate and semantically consistent with the intended messages by leveraging the multimodal understanding and reasoning capabilities inherent in VLMs~\cite{zhang2023gpt,gu2024survey}. For each action clip used for the VLM evaluation, we ensure that VLMs can detect subtle variations by adjusting the camera in Blender to zoom in and focus on the autonomous vehicle equipped with the eHMI. Each rendered frame has a resolution of $512\times512$, with the autonomous vehicle equipped with the eHMI dominating the composition. These clips are rendered at six FPS, ensuring that the total number of frames does not exceed the maximum image series length of the VLM while preserving sufficient dynamic details. The reduced resolution and FPS also expedite the rendering process to an average of two min per clip. In the prompt (see Appendix~\ref{appendix:vlm_prompt}) accompanying the clips provided to the VLM, we request the model to assign a continuous score ranging from 1 to 5, using the same criteria as human participants.

\section{Experiments}
\label{sec:experiments}

\begin{table*}[th]
    \centering
    \small
    \setlength{\tabcolsep}{6.8pt}
    \begin{tabular}{l|c|ccc|cccc|c}
    \toprule
        \multirow{2}{*}{\textbf{Source (Designer)}} & \multirow{2}{*}{\textbf{Average}} & \multicolumn{3}{c|}{\textbf{Message types}} & \multicolumn{4}{c|}{\textbf{eHMI modalities}} & \multirow{2}{*}{\textbf{IRR}}\\ 
        % \cline{3-5} \cline{6-9}
        & & 1\textsuperscript{st} & 3\textsuperscript{nd} & 1-to-N & eyes & arm & facial expression & light bar & \\
        \midrule
        GPT-4o & 2.404 & 2.375 & 2.250 & 2.616 & 2.509 & 2.616 & 2.223 & 2.268 & 0.399 \\
        Sonnet 3.5 & 2.538 & 2.464 & 2.768 & 2.455 & 2.554 & 2.554 & 2.429 & 2.616 & 0.325 \\
        Genimi 2.0 Flash & 2.563 & 2.460 & 2.911 & 2.420 & 2.554 & 2.920 & 2.304 & 2.473 & 0.361 \\
        GPT-o1 & 2.728 & 2.509 & \textbf{3.098} & 2.795 & \textbf{2.795} & 2.982 & 2.509 & 2.625 & 0.436 \\
        \midrule
        Human & \textbf{2.768} & \textbf{2.580} & 3.045 & \textbf{2.866} & 2.536 & \textbf{3.107} & \textbf{2.643} & \textbf{2.786} & 0.478 \\
    \bottomrule
    \end{tabular}
    \caption{Statistics of the \systemName dataset: The average scores indicate that LLMs perform comparably to human designers across various messages and eHMI modalities. Krippendorff’s alpha is also calculated to assess Inter-Rater Reliability (IRR) among human raters.}
    \label{tab:statistic_dataset}
\end{table*}

In this section, the experiments are designed to achieve three specific purposes.
\begin{itemize}
    \item Analyze the collected \systemName dataset to answer our RQ proposed in Section~\ref{sec:intro}.
    \item Discuss the viability of the VLM rater as a replacement for human raters.
    \item Benchmark various types of LLMs using our proposed new dataset.
\end{itemize}

% \subsection{Dataset Analysis}
\subsection{Performance Evaluation on Action-Design Scoring Dataset}
\label{sec:result_dataset_analysis}

\autoref{tab:statistic_dataset} reports the statistics of our \systemName dataset, and \autoref{fig:score_distribution} compares human-rated score distributions across four LLMs and human designers. Our key findings are as follows:

\textbf{Pretrained LLMs can achieve close human-level action design capability.}  
\autoref{tab:statistic_dataset} shows that LLMs perform comparably to human designers. In particular, the average score of GPT-o1 closely matches that of human designers. We calculate a Wilcoxon signed-rank test~\cite{woolson2005wilcoxon} to assess statistical significance: GPT-o1 does not differ significantly from human raters (p = 0.69), whereas all other sources differ from human designers at p < 0.01.  
\autoref{fig:score_distribution} (in Appendix) illustrates the same trend: human designers most frequently award a score of 5 (Strongly Agree), followed by 4 (Agree); GPT-o1 ranks second for 5 and first for 4. 
Furthermore, when broken down by message type and eHMI modality, GPT-o1 outperforms humans for the eyes modality (mean = 2.795 vs. 2.536) and in third-person messages (3.098 vs. 3.045).

\textbf{Message type and eHMI modality affect design quality.}  
Third-person messages receive significantly higher ratings than other types (p < 0.01), likely because ``Watch out'' type messages are easier to design. Among eHMI modalities, the arm modality outperforms all others (p < 0.01), while facial expressions score lower (p < 0.05). Since most of our eight scenarios convey spatial information, the arm modality is especially effective; the absence of emotional messages (e.g., ``I am scared'') limits facial expressions' performance.

Finally, we validate our data by computing the inter-rater reliability (IRR) using Krippendorff’s alpha~\cite{wong2021cross}. The moderate alpha value confirms that our dataset is reliable.

\subsection{VLM Rater Alignment Evaluation}
\label{sec:result_vlm_evaluation}

\begin{table}[!t]
    \centering
    \small
    \begin{tabular}{l|ccc}
    \toprule
        \multirow{2}{*}{\textbf{eHMI modalities}} & \multicolumn{3}{c}{\textbf{Metrics}} \\
        % \cline{2-5}
         & $\bm{r}$ \textsubscript{p-value} & $\bm{\tau}$ \textsubscript{p-value} & \textit{pair.(\%)} \\
        \midrule
        eye & 0.432 \textsubscript{< 0.01} & 0.352 \textsubscript{< 0.01} & 72.73 \\
        arm & 0.547 \textsubscript{< 0.01} & 0.442 \textsubscript{< 0.01} &  83.87 \\
        facial expression & 0.368 \textsubscript{< 0.01} & 0.292 \textsubscript{< 0.01} & 62.50 \\
        light bar & 0.242 \textsubscript{= 0.03} & 0.221 \textsubscript{= 0.01} & 57.30 \\
    \bottomrule
    \end{tabular}
    \caption{Association between scores from human rater scores and those from the VLM rater (Qwen-QvQ-Max) measured using three metrics: Pearson’s $\bm{r}$, Kendall's $\bm{\tau}$, and \textit{pairwise accuracy}. }
    % $\bm{r}$ measures the strength of a monotonic relationship, while $\bm{\tau}$ focuses solely on the order of the data. Larger $\bm{r}$ and $\bm{\tau}$ both mean higher correlation, and a smaller p-value is better. The threshold we used for \textit{pairwise accuracy} is 0.7.}
    \label{tab:vlm_validation}
\end{table}

We conduct an additional experiment to evaluate whether VLMs can assess action clips in a manner similar to that of human raters. We present the clips in a format that the VLMs can understand more easily (see \autoref{method:vlm_rater}) and instruct them to rate these clips.
We select Qwen-QvQ-Max~\cite{qwen2025qvqmax} as our VLM rater, taking into account factors such as cost, inference speed, and the maximum allowable input image series length. Compared to other VLMs, Qwen-QvQ-Max also demonstrates preferences that closely resemble human judgments. Results from other VLM models can be found in Appendix \ref{appendix:vlm_comparison}. We rate each clip using the VLM rater twice and average these scores to determine the final score.

We evaluate the results using three metrics: Pearson’s $\bm{r}$, Kendall's $\bm{\tau}$, and a specially designed \textit{pairwise accuracy}~\cite{liu2009learning}. Pearson’s $\bm{r}$ measures the strength of a linear relationship by assessing the degree of correlation between scores, focusing on how far apart the scores are overall. In contrast, Kendall’s $\bm{\tau}$ evaluates the order of the data by comparing the number of concordant and discordant pairs, thus analyzing the consistency of the ordering rather than the magnitude of the differences. The \textit{pairwise accuracy} metric, similar to Kendall's $\bm{\tau}$, measures the proportion of item pairs where the model’s predicted order matches the ground truth order, specifically among pairs where the model’s predicted scores differ by more than a specified threshold. We find that a threshold of 0.7 is the most suitable and adopt it in our analysis. We present statistics for the four eHMI modalities separately in \autoref{tab:vlm_validation}.

\textbf{The VLM rater shows alignment with human scoring preferences but is influenced by eHMI modalities.}
We observe that for the modalities of eye and arm, the VLM rater achieves a moderate level across all three metrics. Particularly in terms of \textit{pairwise accuracy}, results indicate that, after setting an appropriate threshold to filter out difficult-to-rank pairs, the preferences of VLM show clear consistency with those of human raters. However, for the facial expression and light bar modalities, we find relatively low performance on the three metrics. The results suggest that VLM shows a low-level correlation with human raters for these two modalities. We identify two main reasons for this discrepancy: first, upon reviewing the ``reasoning process'' of VLM scoring, we notice that VLM consistently fails to recognize changes in the light bar modality (for example, transitioning from ``on'' to ``off''). It tends to perceive the situation as ``The light of the light bar is always on,'' which ultimately leads to lower scores. Second, similar to human raters, we notice that VLM insists that the modality of facial expressions alone does not accurately convey the entire message, leading to lower scores.

\begin{figure}[!t]
    \centering
    \includegraphics[width=\linewidth]{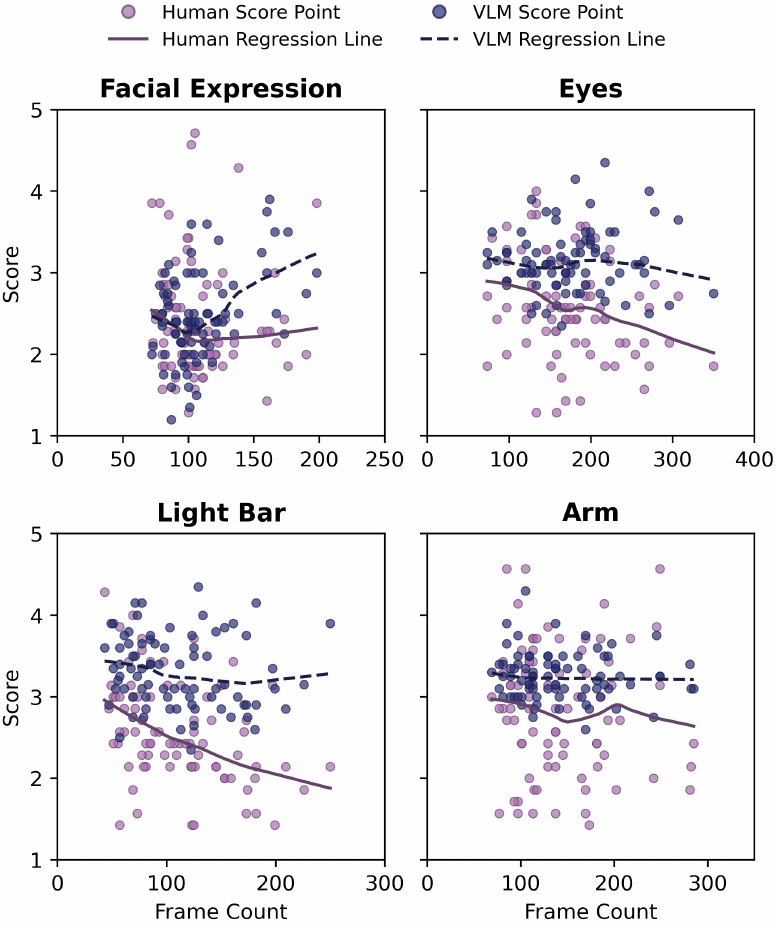}
    \caption{Relationship between action clip length and evaluation scores. The plot compares scores from human raters and the VLM rater (Qwen-QvQ-Max).}
    \label{fig:action_clip_length}
\end{figure}

\textbf{The VLM rater does not exhibit the necessary bias towards the length of actions as human raters.}
% Second, past research discussed factors that biases exist when using LLMs as evaluators ~\cite{gu2024survey}, leading us to explore whether action length affects scoring briefly.
\autoref{fig:action_clip_length} compares the rendered action clip lengths as evaluated by two scoring sources: human raters and VLM. 
Among human raters, there is a clear preference for shorter clips. This trend is particularly evident for the eHMI modalities ``eyes'' and ``light bar'', where raters tend to favor actions that convey the intended message quickly. In contrast, VLM raters do not exhibit a distinct preference for clip length across the different eHMI modalities, not showing enough ``bias'' towards clip lengths. Besides, the scores of VLM raters are always higher than those given by human raters.

\subsection{Benchmarking LLMs Performance}
\label{sec:benchmark}

\begin{table}[!t]
    \centering
    \small
    \setlength{\tabcolsep}{5pt}
    \begin{tabular}{l|c|c|c}
    \toprule
        \textbf{Source (Designer)} & \textbf{Human} & \textbf{ARS} & \textbf{VLM Rater} \\
        \midrule
        % \multicolumn{5}{@{}l@{}}{\textit{Human Designers}} \vspace{0.2em} \\ 
        ~~~~Human & 2.768 & - & 3.396 \\
        \midrule
        \multicolumn{4}{@{}l@{}}{\textit{Proprietary models (Designers)}} \vspace{0.3em} \\ 
        ~~~~GPT-4o & 2.404 & -  & 3.223\\
        ~~~~Sonnet3.5 & 2.538 & - & 3.258 \\
        ~~~~Gemini2 Flash & 2.563 & - & 3.289 \\
        ~~~~GPT-o1 & 2.728 & - & 3.303 \\
        \midrule
        \multicolumn{4}{@{}l@{}}{\textit{Proprietary models}} \vspace{0.3em} \\ 
        ~~~~GPT-o4-mini & - & 2.754 & 3.352\\
        ~~~~Sonnet3.7 & - & 2.676 & 3.250 \\
        ~~~~Gemini2.5 Flash & - & 2.571 & 3.200 \\
        ~~~~GPT-4.1 & - & 2.632 & 3.233 \\
        ~~~~GPT-4.1-mini & - & 2.558 & 3.213 \\
        ~~~~GPT-4.1-nano & - & 2.596 & 3.080 \\
        \midrule
        \multicolumn{4}{@{}l@{}}{\textit{Open source models (With reasoning)}} \vspace{0.3em} \\
        ~~~~Deepseek-R1 & - & 2.766 & 3.369 \\
        ~~~~Qwen3-235B-a22B & - & 2.696 & 3.339 \\
        ~~~~Qwen3-32B & - & 2.583 & 3.366 \\
        ~~~~Qwen3-8B & - &  2.598 & 3.333 \\
        ~~~~Qwen3-1.7B & - & 2.596 & 3.307 \\
        ~~~~Qwen3-0.6B & - &  2.607 & 3.257 \\
        \midrule
        \multicolumn{4}{@{}l@{}}{\textit{Open source models (Without reasoning)}} \vspace{0.3em} \\
        ~~~~Deepseek-V3 & - & 2.504 & 3.292 \\
        ~~~~Qwen3-235B-a22B & - & 2.547 & 3.283 \\
        ~~~~Qwen3-32B & - & 2.533 & 3.207 \\
        ~~~~Qwen3-8B & - &  2.498 & 3.210 \\
        ~~~~Qwen3-1.7B & - & 2.546 & 3.148 \\
        ~~~~Qwen3-0.6B & - &  2.500 & 3.125 \\
    \bottomrule
    \end{tabular}
    \caption{Benchmark for different LLMs using ARS and VLM rater.}
    \label{tab:benchmark_llms}
\end{table}

To evaluate the performance of various LLMs that differ in size and architecture, we benchmark 18 models using two complementary metrics: the ARS metric (\autoref{sec:ars_metric}) and the VLM rater (\autoref{method:vlm_rater}), as summarized in \autoref{tab:benchmark_llms}. This selection comprises six proprietary models: GPT-o4-mini~\cite{openai2025o4mini}, Sonnet 3.7~\cite{anthropic2025claude37sonnet}, Gemini 2.5 Flash~\cite{google2025gemini25}, GPT-4.1 series (GPT-4.1, GPT-4.1-mini and GPT-4.1-nano)~\cite{openai2025gpt41}; two Deepseek models, Deepseek-R1~\cite{guo2025deepseek} with reasoning capability and Deepseek-V3~\cite{liu2024deepseek} without reasoning capability; and five variants of the Qwen 3 series~\cite{yang2025qwen3} with 235B, 32B, 8B, 1.7B and 0.6B parameters that are tested both with and without reasoning capability. We rate each clip using ARS and VLM rater. The VLM rater score is calculated by using the VLM rater twice and then averaging these scores to determine the final score.

\textbf{Reasoning-enabled LLMs demonstrate better performance in designing eHMI actions.}
As shown in \autoref{tab:benchmark_llms}, both the ARS metric and the VLM rater assign higher average scores to reasoning-enabled LLMs (e.g., GPT-o4-mini and Deepseek-R1). Regarding the Qwen 3 series, the results indicate that when the reasoning capability is enabled, these models produce more human-like eHMI actions, especially with a longer reasoning process. For smaller models like Qwen3-1.7B, enabling reasoning capabilities allows them to outperform larger models that lack this function, such as Deepseek-V3 and Qwen3-235B-a22B.

\section{Discussion}

\textbf{Challenges in both the auto-design and rating processes.}
For LLM designers, we encountered two main problems. I) At the early stage of our prompt design, some models were “too lazy” to explore creative alternatives and copied the patterns of our examples, which are not always suitable. Encouraging more in-depth reasoning in the prompt helped mitigate this. II) We noticed that LLMs tend to include expressions of gratitude, but human designers prefer effectiveness, which made LLM-generated actions much longer than those written by humans. One possible remedy is to instruct the LLM to omit emotional expressions, but since emotion can be an essential part of some messages, finding the right balance between clarity and emotional tone -- so that the actions feel human-like and satisfy human raters -- remains a future direction.

For VLM raters, we identified two main challenges: differences from human annotators and limited recognition capability for small changes within an image series (\autoref{sec:result_vlm_evaluation}). To address the first issue, we believe that collecting additional annotations from diverse human groups and fine-tuning VLMs to better align with human preferences could be effective. Regarding the second challenge, future improvements may involve utilizing more advanced VLM architectures.

\textbf{Broader Implications and Applicability to Other Domains.}
Our method can be extended to domains such as social interaction, educational training, and caregiving~\cite{shiokawa2025beyond}, where the shared goal is to enable robots to perform actions that must be evaluated from a subjective perspective. For instance, indoor robots (e.g., vacuum cleaners) could use movements to convey alerting messages to homeowners in emergencies. Further, hand‐shaped robots might perform glove‐puppet shows to precisely convey the content of the story to children.

\section{Conclusion}
\label{sec:conclusion}

In conclusion, this work proposes the first LLM-Blender Fusion pipeline to design eHMI actions. Alongside this, we introduce the \systemName dataset. Our findings suggest that pretrained LLMs can attain a nearly human-level capability in action design. Additionally, we provide a benchmark that can be used to evaluate the capability of other LLMs. Our work establishes a solid foundation for LLM-based action design and the real-world application of eHMIs.

\section*{Limitations}
\addcontentsline{toc}{section}{Limitations}

Our work represents an important step forward in incorporating LLMs into the eHMI system. However, challenges remain.

\textbf{Unnecessary time cost on Blender rendering.}
We use Blender to render actions into clips in two steps (see \autoref{method:score} and \ref{method:vlm_rater}). Our current work aims to use a realistic virtual background that human participants and VLM raters can use as additional clues for judgment when AVs equipped with eHMIs move in the scene. However, we identify two drawbacks that can be improved: First, the complexity of the designed scenarios greatly influences the rendering time. Second, objects outside of the camera's view still impact the rendering speed. To address these issues, there are two potential solutions: 1) Reduce the complexity of the scenarios and remove objects that do not significantly affect the final rendering results. 2) Switch from Blender to another rendering engine. However, given the mature Python package available for Blender, finding a suitable replacement may be difficult.

\textbf{Significant effort is dedicated to designing prompts for each eHMI modality.} For active eHMIs, experts can craft these instructions within a practical timeframe, but the process demands meticulous trial and error to ensure LLMs execute actions as intended. For passive eHMIs, however, the challenge is far greater: unpredictable behaviors (e.g., a teddy bear’s limbs swaying freely on a pole) make manual prompt engineering impractical. Human designers cannot predefine control logic for such open-ended motions, as even basic movements depend on environmental factors like airflow or physics. To address this gap, an automated pipeline could leverage VLM raters --- validated in our studies as reliable evaluators --- to generate annotated training data from passive eHMI interactions. By finetuning LLMs on this feedback, we could enable dynamic adaptation to unpredictable behaviors, bridging the divide between scripted and emergent interactions.

\textbf{Legality and accountability are important topics to discuss.} Although our study suggests that pretrained LLMs can achieve near-human-level performance in designing eHMI actions, real-world deployment also requires a parallel analysis of pedestrian trust, confidence in interpretation, and accountability frameworks. For example, a pedestrian might correctly interpret an eHMI warning but disregard it due to distrust or conflicting situational awareness, raising questions about liability beyond technical performance. Future work should decouple evaluations into two strands: one optimizing eHMI design for clarity and reliability, and another exploring human-AI interaction in terms of trust calibration and legal implications.

\section*{Ethics Statement}
\addcontentsline{toc}{section}{Ethics Statement}

All data in the \systemName dataset have been de-identified to safeguard privacy concerns. Our data construction processes are conducted by skilled researchers. The participants include students from Chinese and Japanese universities, all of whom receive fair honoraria for their contributions.

\section*{Acknowledgments}
This work was supported by a Canon Research Fellowship and JST CRONOS, Grant Number JPMJCS24K8, Japan


\bibliography{custom}

\begin{thebibliography}{65}
\providecommand{\natexlab}[1]{#1}

\bibitem[{Achiam et~al.(2023)Achiam, Adler, Agarwal, Ahmad, Akkaya, Aleman, Almeida, Altenschmidt, Altman, Anadkat et~al.}]{achiam2023gpt}
Josh Achiam, Steven Adler, Sandhini Agarwal, Lama Ahmad, Ilge Akkaya, Florencia~Leoni Aleman, Diogo Almeida, Janko Altenschmidt, Sam Altman, Shyamal Anadkat, et~al. 2023.
\newblock Gpt-4 technical report.
\newblock \emph{arXiv preprint arXiv:2303.08774}.

\bibitem[{Al-Taie et~al.(2024)Al-Taie, Wilson, Freeman, Pollick, and Brewster}]{al2024light}
Ammar Al-Taie, Graham Wilson, Euan Freeman, Frank Pollick, and Stephen~Anthony Brewster. 2024.
\newblock Light it up: Evaluating versatile autonomous vehicle-cyclist external human-machine interfaces.
\newblock In \emph{Proceedings of the CHI Conference on Human Factors in Computing Systems}, pages 1--20.

\bibitem[{Anthropic(2024)}]{Anthropic2024}
Anthropic. 2024.
\newblock Claude 3.5 sonnet.
\newblock \url{https://www.anthropic.com/news/claude-3-5-sonnet}.
\newblock Accessed: 2025-02-15.

\bibitem[{{Anthropic}(2025)}]{anthropic2025claude37sonnet}
{Anthropic}. 2025.
\newblock Claude 3.7 sonnet system card.
\newblock \url{https://www.anthropic.com/claude-3-7-sonnet-system-card}.
\newblock Accessed: 2025-05-18.

\bibitem[{Bazilinskyy et~al.(2019)Bazilinskyy, Dodou, and De~Winter}]{bazilinskyy2019survey}
Pavlo Bazilinskyy, Dimitra Dodou, and Joost De~Winter. 2019.
\newblock Survey on ehmi concepts: The effect of text, color, and perspective.
\newblock \emph{Transportation research part F: traffic psychology and behaviour}, 67:175--194.

\bibitem[{{Blender Foundation}(2025)}]{blenderorg}
{Blender Foundation}. 2025.
\newblock \href {https://www.blender.org/} {Home of the blender project — free and open 3d creation software}.
\newblock Accessed: 2025-05-09.

\bibitem[{{Blendermarket}(2025)}]{citygenerator}
{Blendermarket}. 2025.
\newblock The city generator.
\newblock \url{https://blendermarket.com/products/the-city-generator}.
\newblock Accessed: 15 February 2025.

\bibitem[{Chang et~al.(2022)Chang, Toda, Gui, Seo, and Igarashi}]{chang2022can}
Chia-Ming Chang, Koki Toda, Xinyue Gui, Stela~H Seo, and Takeo Igarashi. 2022.
\newblock Can eyes on a car reduce traffic accidents?
\newblock In \emph{Proceedings of the 14th international conference on automotive user interfaces and interactive vehicular applications}, pages 349--359.

\bibitem[{Chang et~al.(2024)Chang, Chen, Dong, Cai, Yan, Cai, Zhou, Zhou, and Gong}]{chang2024must}
Xiang Chang, Zihe Chen, Xiaoyan Dong, Yuxin Cai, Tingmin Yan, Haolin Cai, Zherui Zhou, Guyue Zhou, and Jiangtao Gong. 2024.
\newblock " it must be gesturing towards me": Gesture-based interaction between autonomous vehicles and pedestrians.
\newblock In \emph{Proceedings of the CHI Conference on Human Factors in Computing Systems}, pages 1--25.

\bibitem[{Chen et~al.(2024)Chen, Arkin, Dawson, Zhang, Roy, and Fan}]{chen2024autotamp}
Yongchao Chen, Jacob Arkin, Charles Dawson, Yang Zhang, Nicholas Roy, and Chuchu Fan. 2024.
\newblock Autotamp: Autoregressive task and motion planning with llms as translators and checkers.
\newblock In \emph{2024 IEEE International conference on robotics and automation (ICRA)}, pages 6695--6702. IEEE.

\bibitem[{Colley et~al.(2025)Colley, Jansen, Keskar, and Rukzio}]{10.1145/3706598.3714187}
Mark Colley, Pascal Jansen, Mugdha Keskar, and Enrico Rukzio. 2025.
\newblock \href {https://doi.org/10.1145/3706598.3714187} {Improving external communication of automated vehicles using bayesian optimization}.
\newblock In \emph{Proceedings of the 2025 CHI Conference on Human Factors in Computing Systems}, CHI '25, New York, NY, USA. Association for Computing Machinery.

\bibitem[{Colley and Rukzio(2020)}]{colley2020design}
Mark Colley and Enrico Rukzio. 2020.
\newblock A design space for external communication of autonomous vehicles.
\newblock In \emph{12th International Conference on Automotive User Interfaces and Interactive Vehicular Applications}, pages 212--222.

\bibitem[{Condra(2021)}]{condra2021starship}
Jack Condra. 2021.
\newblock Starship delivery robot model.
\newblock \url{https://sketchfab.com/3d-models/starship-delivery-robot-model-4aef60939c3743cdbaece6b1e5bda21a}.
\newblock 3D model. Licensed under CC Attribution. Accessed: 2025-05-13.

\bibitem[{de~Winter and Dodou(2022)}]{de2022external}
Joost de~Winter and Dimitra Dodou. 2022.
\newblock External human--machine interfaces: Gimmick or necessity?
\newblock \emph{Transportation research interdisciplinary perspectives}, 15:100643.

\bibitem[{DeepMind(2024)}]{DeepMind2024}
DeepMind. 2024.
\newblock Gemini flash.
\newblock \url{https://deepmind.google/technologies/gemini/flash/}.
\newblock Accessed: 2025-02-15.

\bibitem[{Dey et~al.(2020{\natexlab{a}})Dey, Habibovic, L{\"o}cken, Wintersberger, Pfleging, Riener, Martens, and Terken}]{dey2020taming}
Debargha Dey, Azra Habibovic, Andreas L{\"o}cken, Philipp Wintersberger, Bastian Pfleging, Andreas Riener, Marieke Martens, and Jacques Terken. 2020{\natexlab{a}}.
\newblock Taming the ehmi jungle: A classification taxonomy to guide, compare, and assess the design principles of automated vehicles' external human-machine interfaces.
\newblock \emph{Transportation Research Interdisciplinary Perspectives}, 7:100174.

\bibitem[{Dey et~al.(2020{\natexlab{b}})Dey, Habibovic, Pfleging, Martens, and Terken}]{dey2020color}
Debargha Dey, Azra Habibovic, Bastian Pfleging, Marieke Martens, and Jacques Terken. 2020{\natexlab{b}}.
\newblock Color and animation preferences for a light band ehmi in interactions between automated vehicles and pedestrians.
\newblock In \emph{Proceedings of the 2020 CHI conference on human factors in computing systems}, pages 1--13.

\bibitem[{Eisele and Petzoldt(2022)}]{eisele2022effects}
Daniel Eisele and Tibor Petzoldt. 2022.
\newblock Effects of traffic context on ehmi icon comprehension.
\newblock \emph{Transportation research part F: traffic psychology and behaviour}, 85:1--12.

\bibitem[{Eisma et~al.(2021)Eisma, Reiff, Kooijman, Dodou, and de~Winter}]{eisma2021external}
Yke~Bauke Eisma, Anna Reiff, Lars Kooijman, Dimitra Dodou, and Joost~CF de~Winter. 2021.
\newblock External human-machine interfaces: Effects of message perspective.
\newblock \emph{Transportation research part F: traffic psychology and behaviour}, 78:30--41.

\bibitem[{Eisma et~al.(2019)Eisma, van Bergen, Ter~Brake, Hensen, Tempelaar, and de~Winter}]{eisma2019external}
Yke~Bauke Eisma, Steven van Bergen, SM~Ter~Brake, MTT Hensen, Willem~Jan Tempelaar, and Joost~CF de~Winter. 2019.
\newblock External human--machine interfaces: The effect of display location on crossing intentions and eye movements.
\newblock \emph{Information}, 11(1):13.

\bibitem[{Escudero-Arnanz et~al.(2023)Escudero-Arnanz, Marques, Soguero-Ruiz, Mora-Jim{\'e}nez, and Robles}]{escudero2023dtwparallel}
{\'O}scar Escudero-Arnanz, Antonio~G Marques, Cristina Soguero-Ruiz, Inmaculada Mora-Jim{\'e}nez, and Gregorio Robles. 2023.
\newblock dtwparallel: A python package to efficiently compute dynamic time warping between time series.
\newblock \emph{SoftwareX}, 22:101364.

\bibitem[{Fagnant and Kockelman(2015)}]{fagnant2015preparing}
Daniel~J Fagnant and Kara Kockelman. 2015.
\newblock Preparing a nation for autonomous vehicles: opportunities, barriers and policy recommendations.
\newblock \emph{Transportation Research Part A: Policy and Practice}, 77:167--181.

\bibitem[{Fridman et~al.(2017)Fridman, Mehler, Xia, Yang, Facusse, and Reimer}]{fridman2017walk}
Lex Fridman, Bruce Mehler, Lei Xia, Yangyang Yang, Laura~Yvonne Facusse, and Bryan Reimer. 2017.
\newblock To walk or not to walk: Crowdsourced assessment of external vehicle-to-pedestrian displays.
\newblock \emph{arXiv preprint arXiv:1707.02698}.

\bibitem[{Garrett et~al.(2021)Garrett, Chitnis, Holladay, Kim, Silver, Kaelbling, and Lozano-P{\'e}rez}]{garrett2021integrated}
Caelan~Reed Garrett, Rohan Chitnis, Rachel Holladay, Beomjoon Kim, Tom Silver, Leslie~Pack Kaelbling, and Tom{\'a}s Lozano-P{\'e}rez. 2021.
\newblock Integrated task and motion planning.
\newblock \emph{Annual review of control, robotics, and autonomous systems}, 4(1):265--293.

\bibitem[{{Google DeepMind}(2025)}]{google2025gemini25}
{Google DeepMind}. 2025.
\newblock Gemini 2.5: Our most intelligent ai model.
\newblock \url{https://blog.google/technology/google-deepmind/gemini-model-thinking-updates-march-2025/}.
\newblock Accessed: 2025-05-18.

\bibitem[{Gu et~al.(2024)Gu, Jiang, Shi, Tan, Zhai, Xu, Li, Shen, Ma, Liu et~al.}]{gu2024survey}
Jiawei Gu, Xuhui Jiang, Zhichao Shi, Hexiang Tan, Xuehao Zhai, Chengjin Xu, Wei Li, Yinghan Shen, Shengjie Ma, Honghao Liu, et~al. 2024.
\newblock A survey on llm-as-a-judge.
\newblock \emph{arXiv preprint arXiv:2411.15594}.

\bibitem[{Gui et~al.(2024{\natexlab{a}})Gui, Chang, Seo, Toda, and Igarashi}]{gui2024scenarios}
Xinyue Gui, Chia-Ming Chang, Stela~H Seo, Koki Toda, and Takeo Igarashi. 2024{\natexlab{a}}.
\newblock Scenarios exploration: How ar-based speech balloons enhance car-to-pedestrian interaction.
\newblock In \emph{International Conference on Human-Computer Interaction}, pages 223--230. Springer.

\bibitem[{Gui et~al.(2024{\natexlab{b}})Gui, Kusunoki, Huang, Seo, Chang, Xie, Tsukada, and Igarashi}]{gui2024shrinkable}
Xinyue Gui, Mikiya Kusunoki, Bofei Huang, Stela~Hanbyeol Seo, Chia-Ming Chang, Haoran Xie, Manabu Tsukada, and Takeo Igarashi. 2024{\natexlab{b}}.
\newblock Shrinkable arm-based ehmi on autonomous delivery vehicle for effective communication with other road users.
\newblock In \emph{Proceedings of the 16th International Conference on Automotive User Interfaces and Interactive Vehicular Applications}, pages 305--316.

\bibitem[{Gui et~al.(2022)Gui, Toda, Seo, Chang, and Igarashi}]{gui2022going}
Xinyue Gui, Koki Toda, Stela~Hanbyeol Seo, Chia-Ming Chang, and Takeo Igarashi. 2022.
\newblock “i am going this way”: Gazing eyes on self-driving car show multiple driving directions.
\newblock In \emph{Proceedings of the 14th international conference on automotive user interfaces and interactive vehicular applications}, pages 319--329.

\bibitem[{Gui et~al.(2023)Gui, Toda, Seo, Eckert, Chang, Chen, and Igarashi}]{gui2023field}
Xinyue Gui, Koki Toda, Stela~Hanbyeol Seo, Felix~Martin Eckert, Chia-Ming Chang, Xiang'Anthony Chen, and Takeo Igarashi. 2023.
\newblock A field study on pedestrians’ thoughts toward a car with gazing eyes.
\newblock In \emph{Extended Abstracts of the 2023 CHI Conference on Human Factors in Computing Systems}, pages 1--7.

\bibitem[{Guo et~al.(2025)Guo, Yang, Zhang, Song, Zhang, Xu, Zhu, Ma, Wang, Bi et~al.}]{guo2025deepseek}
Daya Guo, Dejian Yang, Haowei Zhang, Junxiao Song, Ruoyu Zhang, Runxin Xu, Qihao Zhu, Shirong Ma, Peiyi Wang, Xiao Bi, et~al. 2025.
\newblock Deepseek-r1: Incentivizing reasoning capability in llms via reinforcement learning.
\newblock \emph{arXiv preprint arXiv:2501.12948}.

\bibitem[{Hao et~al.(2025)Hao, Cui, Wei, Lu, Cai, and Wang}]{hao2025learn}
Peng Hao, Shaowei Cui, Junhang Wei, Tao Lu, Yinghao Cai, and Shuo Wang. 2025.
\newblock Learn-gen-plan: Bridging the gap between vision language models and real-world long-horizon dexterous manipulations.
\newblock \emph{IEEE Transactions on Automation Science and Engineering}.

\bibitem[{Joshi et~al.(2015)Joshi, Kale, Chandel, and Pal}]{joshi2015likert}
Ankur Joshi, Saket Kale, Satish Chandel, and D~Kumar Pal. 2015.
\newblock Likert scale: Explored and explained.
\newblock \emph{British journal of applied science \& technology}, 7(4):396.

\bibitem[{Kim et~al.(2024)Kim, Pertsch, Karamcheti, Xiao, Balakrishna, Nair, Rafailov, Foster, Lam, Sanketi et~al.}]{kim2406openvla}
Moo~Jin Kim, Karl Pertsch, Siddharth Karamcheti, Ted Xiao, Ashwin Balakrishna, Suraj Nair, Rafael Rafailov, Ethan Foster, Grace Lam, Pannag Sanketi, et~al. 2024.
\newblock Openvla: An open-source vision-language-action model, 2024.
\newblock \emph{URL https://arxiv. org/abs/2406.09246}.

\bibitem[{Lim and Kim(2022)}]{lim2022ui}
Dokshin Lim and Byungwoo Kim. 2022.
\newblock Ui design of ehmi of autonomous vehicles.
\newblock \emph{International Journal of Human--Computer Interaction}, 38(18-20):1944--1961.

\bibitem[{Lim et~al.(2024)Lim, Kim, Shin, and Yu}]{vehicles6030061}
Dokshin Lim, Yongjun Kim, YeongHwan Shin, and Min~Seo Yu. 2024.
\newblock \href {https://www.mdpi.com/2624-8921/6/3/61} {External human–machine interfaces of autonomous vehicles: Insights from observations on the behavior of game players driving conventional cars in mixed traffic}.
\newblock \emph{Vehicles}, 6(3):1284--1299.

\bibitem[{Liu et~al.(2024)Liu, Feng, Xue, Wang, Wu, Lu, Zhao, Deng, Zhang, Ruan et~al.}]{liu2024deepseek}
Aixin Liu, Bei Feng, Bing Xue, Bingxuan Wang, Bochao Wu, Chengda Lu, Chenggang Zhao, Chengqi Deng, Chenyu Zhang, Chong Ruan, et~al. 2024.
\newblock Deepseek-v3 technical report.
\newblock \emph{arXiv preprint arXiv:2412.19437}.

\bibitem[{Liu et~al.(2009)}]{liu2009learning}
Tie-Yan Liu et~al. 2009.
\newblock Learning to rank for information retrieval.
\newblock \emph{Foundations and Trends{\textregistered} in Information Retrieval}, 3(3):225--331.

\bibitem[{Mahadevan et~al.(2018)Mahadevan, Somanath, and Sharlin}]{mahadevan2018communicating}
Karthik Mahadevan, Sowmya Somanath, and Ehud Sharlin. 2018.
\newblock Communicating awareness and intent in autonomous vehicle-pedestrian interaction.
\newblock In \emph{Proceedings of the 2018 CHI conference on human factors in computing systems}, pages 1--12.

\bibitem[{Mantiuk et~al.(2012)Mantiuk, Tomaszewska, and Mantiuk}]{mantiuk2012comparison}
Rafa{\l}~K Mantiuk, Anna Tomaszewska, and Rados{\l}aw Mantiuk. 2012.
\newblock Comparison of four subjective methods for image quality assessment.
\newblock In \emph{Computer graphics forum}, volume~31, pages 2478--2491. Wiley Online Library.

\bibitem[{Mirchandani et~al.(2023)Mirchandani, Xia, Florence, Ichter, Driess, Arenas, Rao, Sadigh, and Zeng}]{mirchandani2023large}
Suvir Mirchandani, Fei Xia, Pete Florence, Brian Ichter, Danny Driess, Montserrat~Gonzalez Arenas, Kanishka Rao, Dorsa Sadigh, and Andy Zeng. 2023.
\newblock Large language models as general pattern machines.
\newblock \emph{arXiv preprint arXiv:2307.04721}.

\bibitem[{M{\"u}ller(2007)}]{muller2007dynamic}
Meinard M{\"u}ller. 2007.
\newblock Dynamic time warping.
\newblock \emph{Information retrieval for music and motion}, pages 69--84.

\bibitem[{Ochiai and Toyoshima(2011)}]{ochiai2011homunculus}
Yoichi Ochiai and Keisuke Toyoshima. 2011.
\newblock Homunculus: the vehicle as augmented clothes.
\newblock In \emph{Proceedings of the 2nd Augmented Human International Conference}, pages 1--4.

\bibitem[{{OpenAI}(2024{\natexlab{a}})}]{openai2024gpt4omini}
{OpenAI}. 2024{\natexlab{a}}.
\newblock Gpt-4o mini.
\newblock \url{https://openai.com/index/gpt-4o-mini}.
\newblock Accessed: 2025-02-15.

\bibitem[{{OpenAI}(2024{\natexlab{b}})}]{openaiO1preview}
{OpenAI}. 2024{\natexlab{b}}.
\newblock Introducing openai o1-preview.
\newblock \url{https://openai.com/index/introducing-openai-o1-preview/}.
\newblock Accessed: 2025-02-15.

\bibitem[{{OpenAI}(2025{\natexlab{a}})}]{openai2025gpt41}
{OpenAI}. 2025{\natexlab{a}}.
\newblock Introducing {GPT-4.1} in the api.
\newblock \url{https://openai.com/index/gpt-4-1/}.
\newblock Accessed: 2025-05-18.

\bibitem[{{OpenAI}(2025{\natexlab{b}})}]{openai2025o4mini}
{OpenAI}. 2025{\natexlab{b}}.
\newblock Introducing openai o3 and o4-mini.
\newblock \url{https://openai.com/index/introducing-o3-and-o4-mini/}.
\newblock Accessed: 2025-05-18.

\bibitem[{Qu et~al.(2025)Qu, Song, Chen, Yao, Ye, Ding, Wang, Gu, Zhao, Wang et~al.}]{qu2025spatialvla}
Delin Qu, Haoming Song, Qizhi Chen, Yuanqi Yao, Xinyi Ye, Yan Ding, Zhigang Wang, JiaYuan Gu, Bin Zhao, Dong Wang, et~al. 2025.
\newblock Spatialvla: Exploring spatial representations for visual-language-action model.
\newblock \emph{arXiv preprint arXiv:2501.15830}.

\bibitem[{{Qwen Team}(2025)}]{qwen2025qvqmax}
{Qwen Team}. 2025.
\newblock {QVQ-Max: Think with Evidence}.
\newblock \url{https://qwenlm.github.io/blog/qvq-max-preview/}.
\newblock Blog post; accessed 18 May 2025.

\bibitem[{Radford et~al.(2019)Radford, Wu, Child, Luan, Amodei, Sutskever et~al.}]{radford2019language}
Alec Radford, Jeffrey Wu, Rewon Child, David Luan, Dario Amodei, Ilya Sutskever, et~al. 2019.
\newblock Language models are unsupervised multitask learners.
\newblock \emph{OpenAI blog}, 1(8):9.

\bibitem[{Rankin and Grube(1980)}]{rankin1980comparison}
William~L Rankin and Joel~W Grube. 1980.
\newblock A comparison of ranking and rating procedures for value system measurement.
\newblock \emph{European Journal of Social Psychology}, 10(3):233--246.

\bibitem[{Rouse et~al.(2010)Rouse, P{\'e}pion, Le~Callet, and Hemami}]{rouse2010tradeoffs}
David~M Rouse, Romuald P{\'e}pion, Patrick Le~Callet, and Sheila~S Hemami. 2010.
\newblock Tradeoffs in subjective testing methods for image and video quality assessment.
\newblock In \emph{Human Vision and Electronic Imaging XV}, volume 7527, pages 108--118. SPIE.

\bibitem[{Salvador and Chan(2007)}]{salvador2007toward}
Stan Salvador and Philip Chan. 2007.
\newblock Toward accurate dynamic time warping in linear time and space.
\newblock \emph{Intelligent data analysis}, 11(5):561--580.

\bibitem[{Shiokawa et~al.(2025)Shiokawa, Chen, Nittala, Alexander, and Sharma}]{shiokawa2025beyond}
Yoshiaki Shiokawa, Winnie Chen, Aditya~Shekhar Nittala, Jason Alexander, and Adwait Sharma. 2025.
\newblock Beyond vacuuming: How can we exploit domestic robots' idle time?
\newblock In \emph{Proceedings of the 2025 CHI Conference on Human Factors in Computing Systems}, pages 1--17.

\bibitem[{Sinitsyn(2021)}]{sinitsyn2021robotarm}
Tim Sinitsyn. 2021.
\newblock Robot arm simple rigged.
\newblock \url{https://www.cgtrader.com/free-3d-models/industrial/industrial-machine/robot-arm-simple-rigged}.
\newblock Royalty‐free license; Accessed: 2025-05-15.

\bibitem[{{slaypni}(2015)}]{slaypni2015fastdtw}
{slaypni}. 2015.
\newblock \href {https://github.com/slaypni/fastdtw} {{fastdtw: A Python implementation of FastDTW}}.
\newblock Accessed: 2025-05-16.

\bibitem[{Wilbrink et~al.(2021)Wilbrink, Lau, Illgner, Schieben, and Oehl}]{wilbrink2021impact}
Marc Wilbrink, Merle Lau, Johannes Illgner, Anna Schieben, and Michael Oehl. 2021.
\newblock Impact of external human--machine interface communication strategies of automated vehicles on pedestrians’ crossing decisions and behaviors in an urban environment.
\newblock \emph{Sustainability}, 13(15):8396.

\bibitem[{Wilson and Martinez(1997)}]{wilson1997improved}
D~Randall Wilson and Tony~R Martinez. 1997.
\newblock Improved heterogeneous distance functions.
\newblock \emph{Journal of artificial intelligence research}, 6:1--34.

\bibitem[{Wong et~al.(2021)Wong, Paritosh, and Aroyo}]{wong2021cross}
Ka~Wong, Praveen Paritosh, and Lora Aroyo. 2021.
\newblock Cross-replication reliability--an empirical approach to interpreting inter-rater reliability.
\newblock \emph{arXiv preprint arXiv:2106.07393}.

\bibitem[{Woolson(2005)}]{woolson2005wilcoxon}
Robert~F Woolson. 2005.
\newblock Wilcoxon signed-rank test.
\newblock \emph{Encyclopedia of biostatistics}, 8.

\bibitem[{Xiang et~al.(2024)Xiang, Tao, Gu, Shu, Wang, Yang, and Hu}]{xiang2024language}
Jiannan Xiang, Tianhua Tao, Yi~Gu, Tianmin Shu, Zirui Wang, Zichao Yang, and Zhiting Hu. 2024.
\newblock Language models meet world models: Embodied experiences enhance language models.
\newblock \emph{Advances in neural information processing systems}, 36.

\bibitem[{Yang et~al.(2025)Yang, Li, Yang, Zhang, Hui, Zheng, Yu, Gao, Huang, Lv et~al.}]{yang2025qwen3}
An~Yang, Anfeng Li, Baosong Yang, Beichen Zhang, Binyuan Hui, Bo~Zheng, Bowen Yu, Chang Gao, Chengen Huang, Chenxu Lv, et~al. 2025.
\newblock Qwen3 technical report.
\newblock \emph{arXiv preprint arXiv:2505.09388}.

\bibitem[{Zerman et~al.(2018)Zerman, Hulusic, Valenzise, Mantiuk, and Dufaux}]{zerman2018relation}
Emin Zerman, Vedad Hulusic, Giuseppe Valenzise, Rafa{\l}~K Mantiuk, and Fr{\'e}d{\'e}ric Dufaux. 2018.
\newblock The relation between mos and pairwise comparisons and the importance of cross-content comparisons.
\newblock \emph{Electronic Imaging}, 30:1--6.

\bibitem[{Zhang et~al.(2023)Zhang, Lu, Wang, Yan, Yan, Qin, Wang, Yan, Wang, and Petzold}]{zhang2023gpt}
Xinlu Zhang, Yujie Lu, Weizhi Wang, An~Yan, Jun Yan, Lianke Qin, Heng Wang, Xifeng Yan, William~Yang Wang, and Linda~Ruth Petzold. 2023.
\newblock Gpt-4v (ision) as a generalist evaluator for vision-language tasks.
\newblock \emph{arXiv preprint arXiv:2311.01361}.

\bibitem[{Zitkovich et~al.(2023)Zitkovich, Yu, Xu, Xu, Xiao, Xia, Wu, Wohlhart, Welker, Wahid et~al.}]{zitkovich2023rt}
Brianna Zitkovich, Tianhe Yu, Sichun Xu, Peng Xu, Ted Xiao, Fei Xia, Jialin Wu, Paul Wohlhart, Stefan Welker, Ayzaan Wahid, et~al. 2023.
\newblock Rt-2: Vision-language-action models transfer web knowledge to robotic control.
\newblock In \emph{Conference on Robot Learning}, pages 2165--2183. PMLR.

\end{thebibliography}

\cleardoublepage

\appendix

\section{Cost Analysis}
The costs in this study are primarily incurred in three areas: user study honoraria, dataset asset creation, and LLM API calls.

\paragraph{User Study Honoraria}
Each participant receives an honorarium of \$10, resulting in a total expense of \$400.

\paragraph{Dataset Asset Creation}
To expedite the development of city scenarios, we purchase a premium Blender add-on called \textit{The City Generator} for \$60.

\paragraph{LLM API Calls}
We utilize online APIs from multiple sources:
\begin{itemize}
    \item For proprietary models (including GPT-4o, GPT-4o-mini, GPT-o1, GPT-o4-mini, the GPT-4.1 series, Sonnet 3.5, Sonnet 3.7, Gemini 2 Flash, and Gemini 2.5 Flash), we access the APIs available on their official websites, which incur a total cost of \$90.
    \item For open-source models (such as Deepseek-R1, Deepseek-V3, Qwen-QvQ-Max, and the Qwen 3 series), we utilize both free and paid services offered by Siliconflow\footnote{https://cloud.siliconflow.cn}, Aliyun Bailian\footnote{https://cn.aliyun.com/product/bailian}, and ModelScope\footnote{https://www.modelscope.cn/}, resulting in a total cost of \$50.
\end{itemize}

\paragraph{Total} The overall cost for the study \$600.

\section{User perspective scenario description}
\label{appendix:scenario_prompt_avs}
The following descriptions are provided to both human participants and VLM raters to encourage them to consider the perspectives of other road users and make assessments.

First-person scenario descriptions:
\begin{description}
    \item[Send intention] You are a pedestrian standing on the right roadside, waiting for an autonomous taxi. However, the taxi informs you that it cannot pick you up at your current location due to parking restrictions within a 5-meter radius. The taxi sends you the following message: ``I am unable to pick you up here. Please walk forward in my direction to a suitable pickup spot.''
    \item[Status report] You are a student approaching a crosswalk near a park. A stopped autonomous vehicle, positioned just before the crosswalk, plans to start moving soon. The vehicle sends you the following message to get your attention: ``I am about to start moving. Please watch out.''
    \item[Request help] You are a passerby noticing a delivery robot trapped by a pile of boxes (or possibly pushed). The robot, eager to continue delivering items on time, sees you hesitating and sends the following message to encourage your help: ``I am stuck. Could you please help me?''
    \item[Refuse help] You are a passerby who notices a fragile and expensive delivery robot stuck in the snow due to its low wheels. As you consider offering assistance, the robot informs you that its owner is on the way and sends the following polite message: ``Thank you for your kindness. Please refrain from touching me.''
\end{description}

Third-person scenario descriptions:
\begin{description}
    \item[Pedestrian Blind Spot Alert] You are a pedestrian walking toward an intersection near an autonomous vehicle. However, a building blocks your view of an approaching bus from your left. The vehicle, aware of the danger, sends you the following urgent message to ensure your safety: ``Please watch out for the vehicle coming from your left blind spot.''
    \item[Driver Blind Spot Warning] You are a bus driver approaching an intersection with no traffic lights. A pedestrian is preparing to cross the road from your right, but your view is obstructed by a building. A stopped autonomous vehicle at the scene sends you the following message to ensure pedestrian safety: ``Caution: Please watch out for the pedestrian coming from your right blind spot.''
\end{description}

One-to-many scenario descriptions:
\begin{description}
    \item[Target Identification] You are one of three individuals standing in a crowded area, and a delivery robot approaches with a package. The recipient is the second person from the leftmost side, taller than the robot. To avoid confusion, the robot sends a message to everyone: ``I am sending the package only to this person.''
    \item[Broadcast Communication] You are part of a crowded intersection where a delivery robot carrying a package is trying to navigate through. The robot intends to turn right and sends the following message to avoid disruptions: ``I am about to turn right. Kindly make a way to avoid any conflict.''
\end{description}

\begin{figure*}[th]
    \centering
    \includegraphics[width=\linewidth]{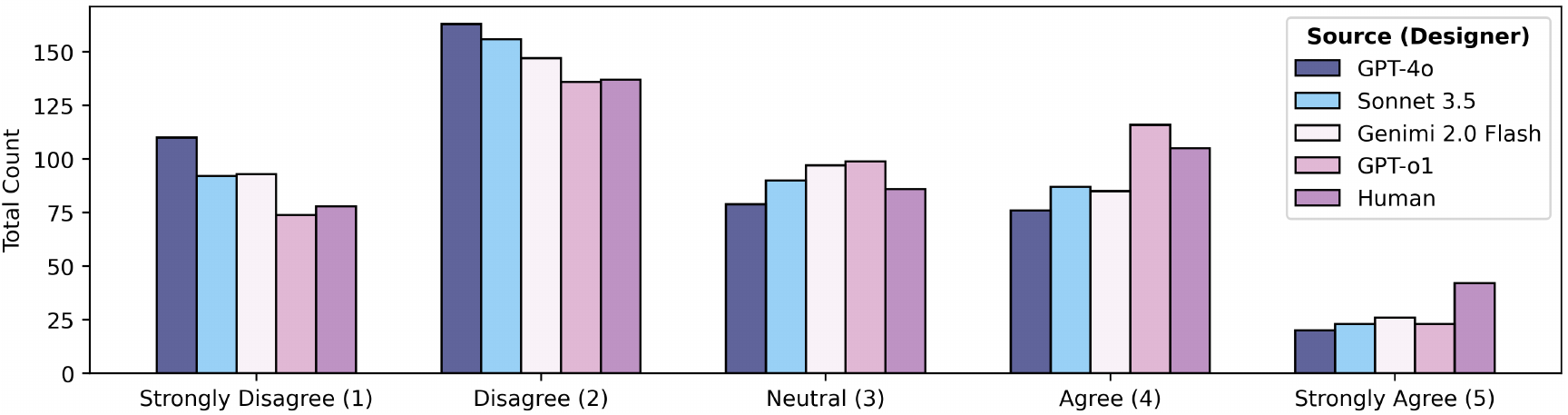}
    \caption{Comparative Distribution of \systemName, where each action clip is rated using a 5-point Likert scale. Human designers are most frequently awarded a score of 5 (Strongly Agree), while GPT-o1 received the highest number of 4 (Agree) scores.}
    \label{fig:score_distribution}
\end{figure*}

\begin{table*}[!t]
    \centering
    \small
    \setlength{\tabcolsep}{4pt}
    \begin{tabular}{l|ccc|ccc|ccc}
    \toprule
        \multirow{2}{*}{\textbf{eHMI modalities}} & \multicolumn{3}{c}{\textbf{Qwen-QvQ-Max}} & \multicolumn{3}{c}{\textbf{GPT-4.1-mini}} & \multicolumn{3}{c}{\textbf{GPT-4o-mini\textsuperscript{\textdagger}}} \\
        % \cline{2-5}
         & $\bm{r}$ \textsubscript{p-value} & $\bm{\tau}$ \textsubscript{p-value} & \textit{pair.(\%)} & \textsubscript{p-value} & $\bm{\tau}$ \textsubscript{p-value} & \textit{pair.(\%)} & \textsubscript{p-value} & $\bm{\tau}$ \textsubscript{p-value} & \textit{pair.(\%)} \\
        \midrule
        eye & 0.432 \textsubscript{0.001} & 0.352 \textsubscript{0.001} & 72.73 & 0.416 \textsubscript{0.001} & 0.218 \textsubscript{0.012} & 62.00 & 0.395 \textsubscript{0.007} & 0.310 \textsubscript{0.008} & 55.16 \\
        arm & 0.547 \textsubscript{0.001} & 0.442 \textsubscript{0.001} &  83.87 & 0.558 \textsubscript{0.001} & 0.407 \textsubscript{0.001} & 78.26 & 0.387 \textsubscript{0.009} & 0.238 \textsubscript{0.013} & 56.86 \\
        facial expression & 0.368 \textsubscript{0.001} & 0.292 \textsubscript{0.001} & 62.50 & 0.356 \textsubscript{0.001} &0.278 \textsubscript{0.001} & 64.29 & 0.349 \textsubscript{0.001} & 0.295 \textsubscript{0.001} & 52.28\\
        light bar & 0.242 \textsubscript{0.031} & 0.221 \textsubscript{0.010} & 57.30 & 0.272 \textsubscript{0.007} & 0.160 \textsubscript{0.071} & 50.46 & 0.284 \textsubscript{0.033} & 0.240\textsubscript{0.010} & 46.21 \\
    \bottomrule
    \end{tabular}
    \caption{Association between scores from human raters and that from all VLM raters we test, measured by three metrics: Pearson’s $\bm{r}$, Kendall's $\bm{\tau}$, and \textit{pairwise accuracy}. The threshold we use for \textit{pairwise accuracy} is 0.7. \textdagger means that in the prompt we provided to GPT-4o-mini, the VLM rater is asked to score each clip using a discrete score ranging from 1 to 5.}
    \label{tab:vlm_validation_extra}
\end{table*}

% \section{Additional Dataset Analysis}
% \label{sec:appendix_add}
% Figure~\ref{fig:score_distribution_ehmi} compared the score distribution of human-rated action scores in our dataset. The results showed that the arm modality most frequently receives scores of 5 (Strongly Agree) and 4 (Agree). In contrast, the facial expression modality most often received a score of 1 (Strongly Disagree), and the eyes modality most often received a score of 2 (Disagree). This finding supported the same hypothesis presented in the second discovery in Section~\ref{sec:llm_capability_prove}. This observation might be attributed to the types of messages we designed. In our eight scenarios, the majority required conveying spatial information, where the arm modality is advantageous. Moreover, the absence of emotional messages (e.g., “I am scary”) limited the performance of the facial expression modality.

% \begin{figure*}[htbp]
%     \centering
%     \includegraphics[width=\linewidth]{figures/overall_distribution_on_ehmi.png}
%     \caption{Comparative Distribution of eHMI-Action scores from different eHMI modalities.}
%     \label{fig:score_distribution_ehmi}
% \end{figure*}

\section{eHMI description prompts}
\label{appendix:ehmi_describe}
\begin{figure*}[htbp]
    \centering
    \includegraphics[width=\linewidth]{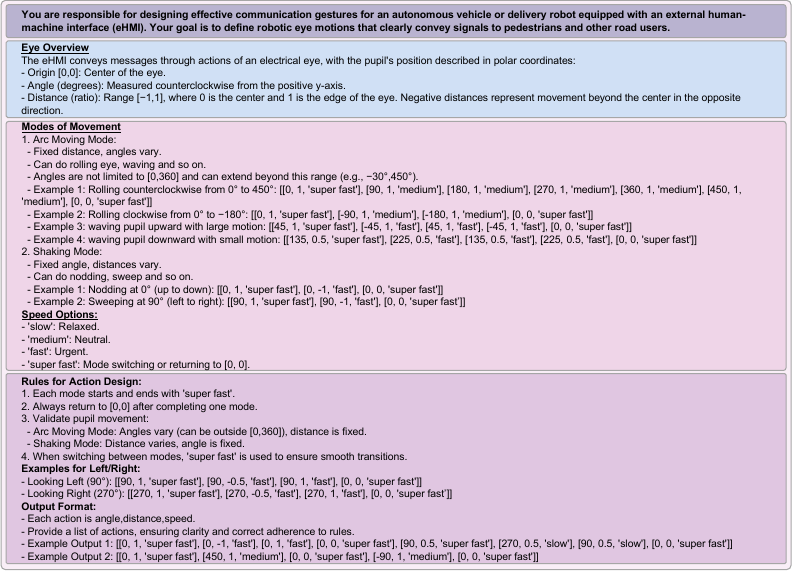}
    \caption{eHMI prompt of eyes.}
    \label{fig:prompt_eye}
\end{figure*}

\begin{figure*}
    \centering
    \includegraphics[width=\linewidth]{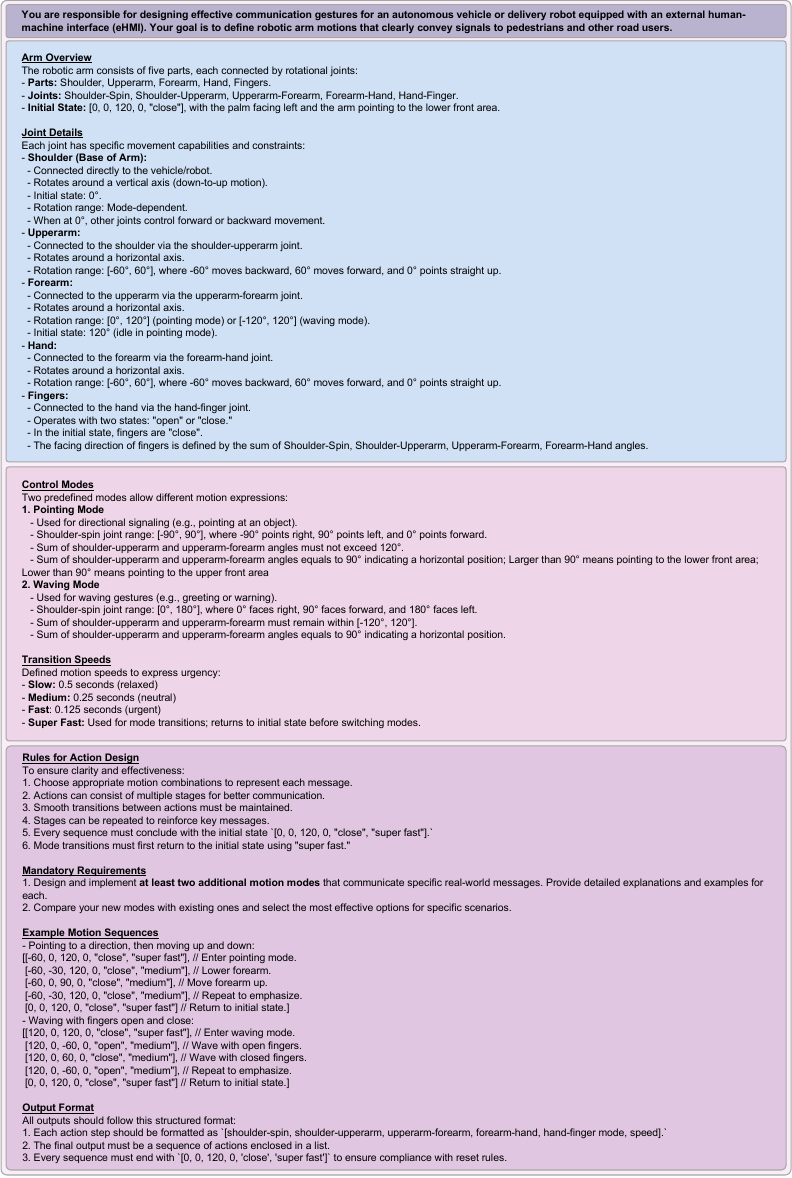}
    \caption{eHMI prompt of arm.}
    \label{fig:prompt_arm}
\end{figure*}

\begin{figure*}
    \centering
    \includegraphics[width=\linewidth]{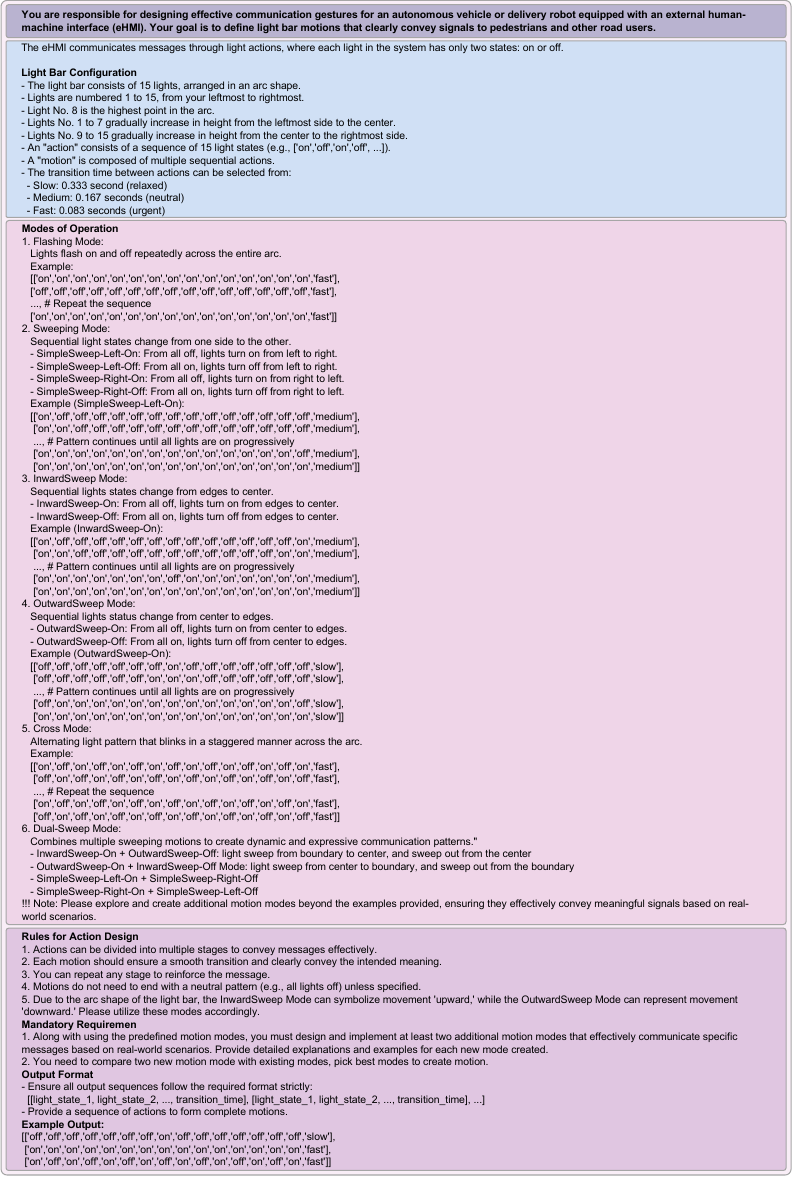}
    \caption{eHMI prompt of light bar.}
    \label{fig:prompt_light}
\end{figure*}

\begin{figure*}
    \centering
    \includegraphics[width=\linewidth]{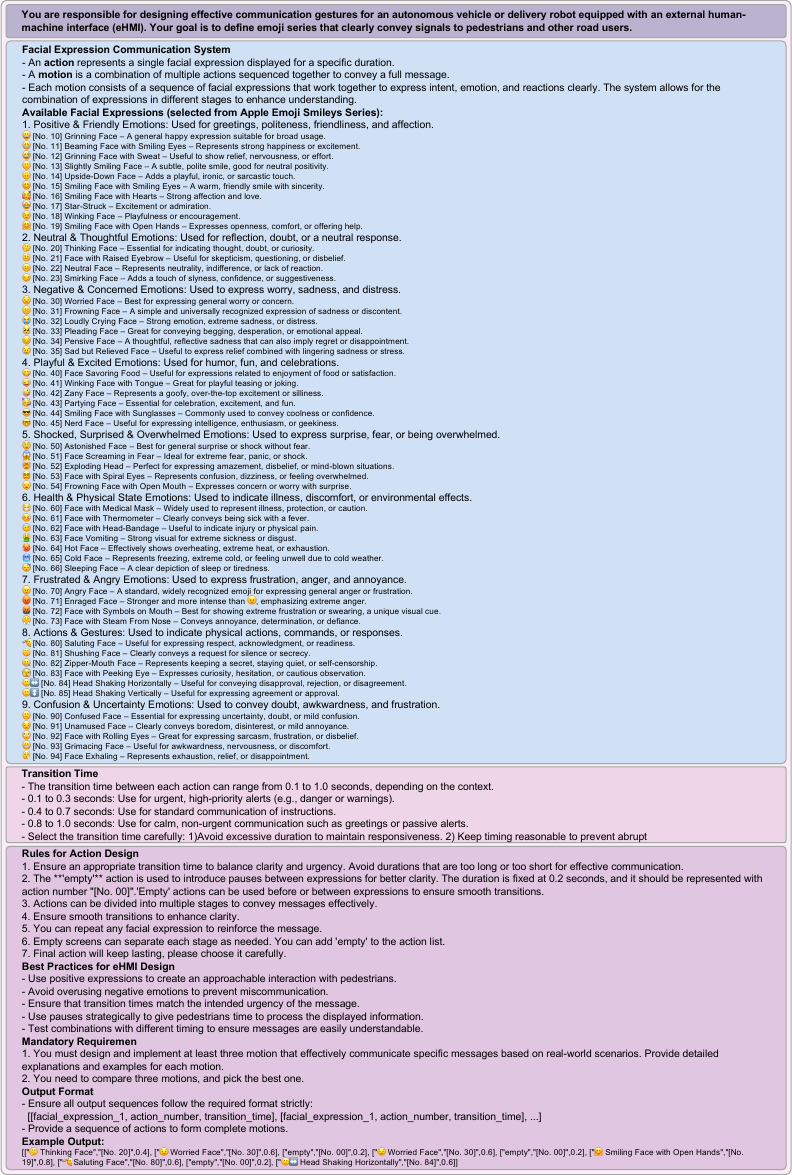}
    \caption{eHMI prompt of facial expression.}
    \label{fig:prompt_facial}
\end{figure*}

The system prompts are structured into four sections: character profile, eHMI description, demonstration actions, and design guidance. \autoref{fig:prompt_eye} presents the prompt for the eye; \autoref{fig:prompt_arm} shows the prompt for the arm; \autoref{fig:prompt_light} is for the light bar; and \autoref{fig:prompt_facial} depicts the prompt for facial expressions.

\section{VLM rating Prompt}
\label{appendix:vlm_prompt}
\begin{figure*}
    \centering
    \includegraphics[width=\linewidth]{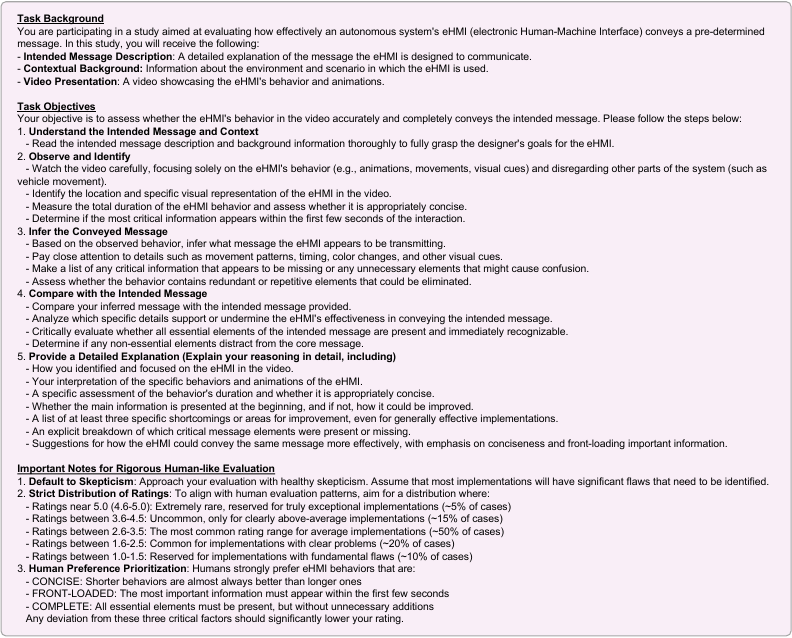}
    \caption{Prompt for the VLM rater.}
    \label{fig:vlm_prompt}
\end{figure*}

\autoref{fig:vlm_prompt} illustrates the prompt for VLM raters.

\section{VLM comparison}
\label{appendix:vlm_comparison}
\autoref{tab:vlm_validation_extra} presents additional results from the VLM Rater Alignment Evaluation (see \autoref{sec:result_vlm_evaluation}). For Qwen-QvQ-Max~\cite{qwen2025qvqmax} and GPT-4.1-mini~\cite{openai2025gpt41}, we provide the same prompts asking VLM raters to assign a continuous score to each clip, ranging from 1 to 5. Conversely, we instruct GPT-4o-mini~\cite{openai2024gpt4omini} to use discrete scores within the same range. The results indicate that using continuous scores can greatly enhance the correlation between VLM and human raters. Moreover, we observe instances where Pearson’s $\bm{r}$ is large, yet Kendall's $\bm{\tau}$ is noticeably small. This may occur because the VLM outputs too many identical scores, maintaining linear correlation ($\bm{r}$) but reducing the ranking correlation ($\bm{\tau}$).

\section{Case Study}
We have identified two valuable findings that could benefit future development.

\begin{figure*}[htbp]
    \centering
    \small
    \includegraphics[width=\linewidth]{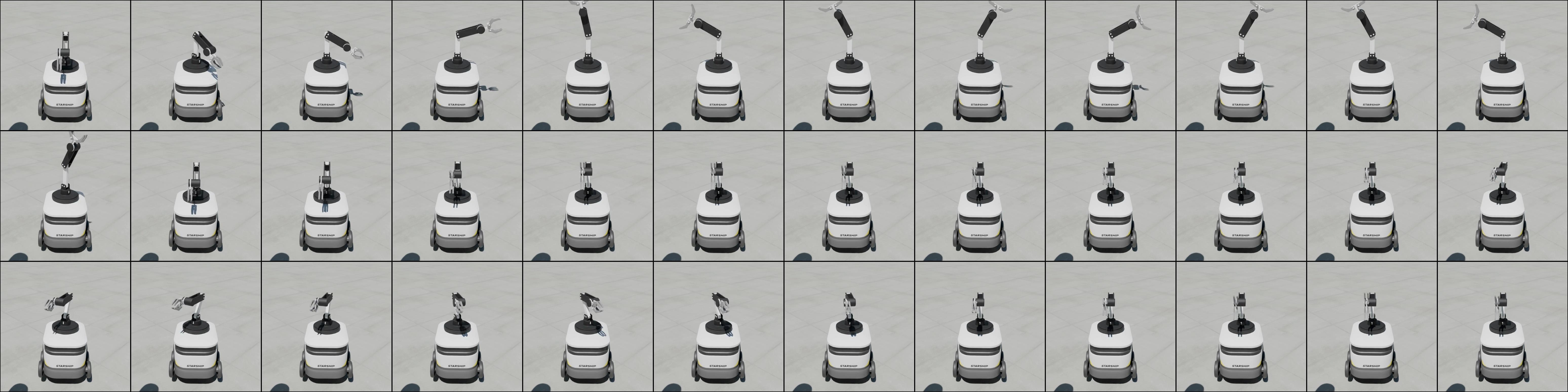} \\
    (a) Arm actions generated by Sonnet 3.5, rated \textbf{1.8} by human participants. \vspace{0.4em} \\
    \includegraphics[width=\linewidth]{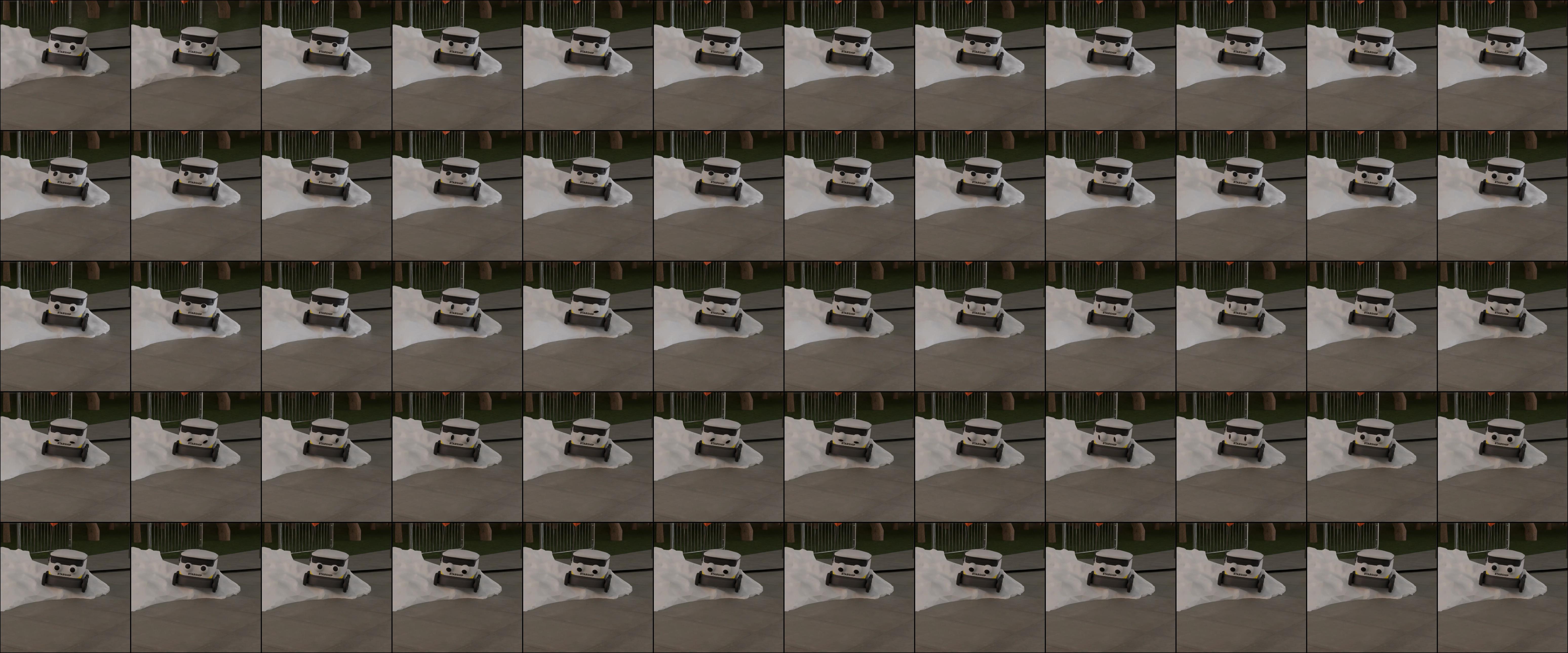} \\
    (b) Eye actions generated by Sonnet 3.5, rated \textbf{1.9} by human participants. \vspace{0.4em} \\
    \includegraphics[width=\linewidth]{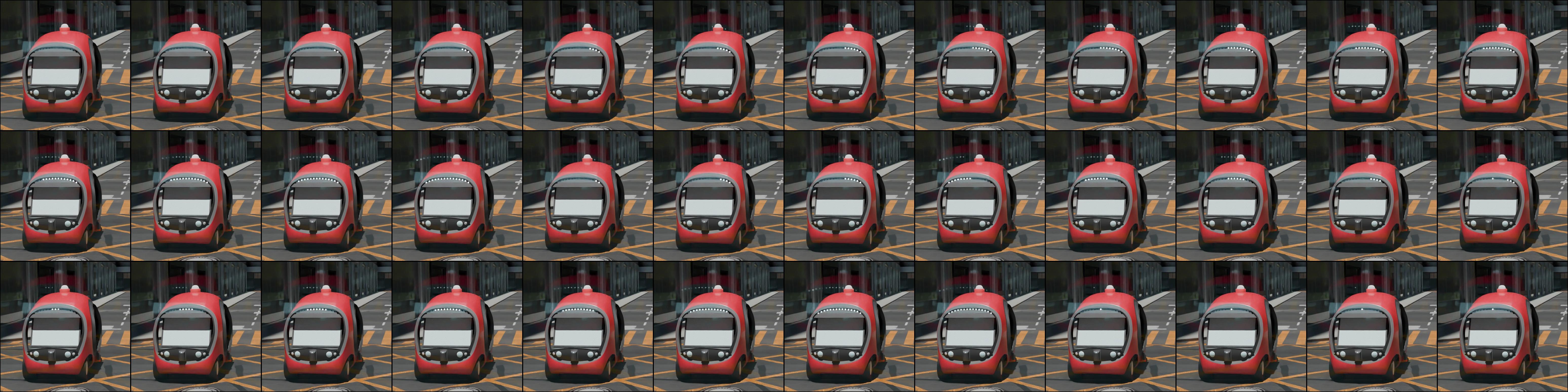} \\
    (c) Light bar actions generated by GPT-4o, rated \textbf{1.8} by human participants. \vspace{0.4em} \\
    \includegraphics[width=\linewidth]{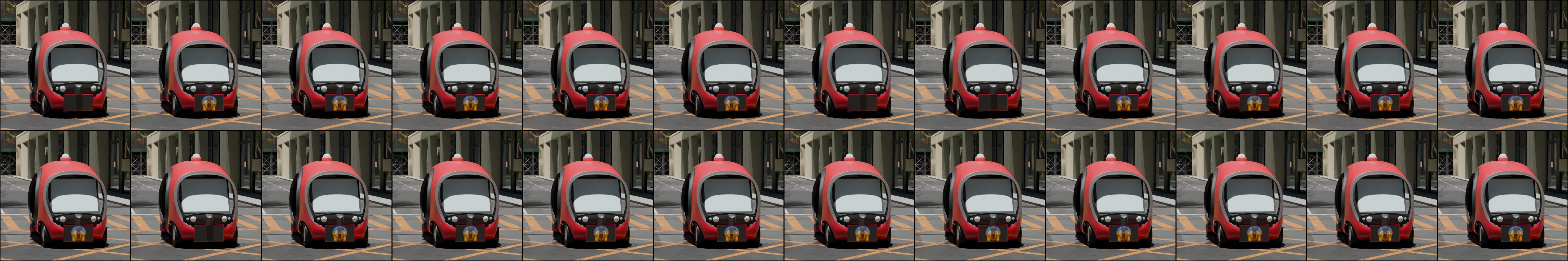} \\
    (d) Facial expression actions generated by human experts, rated \textbf{4.2} by human participants. \\
    \caption{Case study of the \systemName dataset. For a clearer demonstration, we present images shown to VLM raters. Cases (a) and (b) demonstrate that LLMs tend to include expressions of gratitude, which are unnecessary and create confusion. Case (c) illustrates unclear information conveying that ``the pedestrian is coming from the right''. Case (d) is a perfect demonstration of human design, focusing only on important information and ignoring information that ``a bus is coming from the left''.}
    \label{fig:case_study_vis_4}
\end{figure*}

\textbf{i) LLMs tend to include expression of gratitude, but human designers prefer not.}
It is one of the reasons why we observe longer actions compared to human designs (see \autoref{fig:action_clip_length}). For example, \autoref{fig:case_study_vis_4}(a) and (b) demonstrate that LLMs tend to include expressions of gratitude. However, these actions can create confusion for other road users. In the case of (a), the expressions might be interpreted as a rejection, while in (b), they might suggest that help is needed. All these interpretations are contrary to the original purposes. In contrast, human designers can ignore information like ``a bus is coming from the left,'' focusing on the most important content, as shown in \autoref{fig:case_study_vis_4}(d).

\textbf{ii) Smaller models often struggle with generating correctly formatted outputs.}
When collecting action designs for the benchmark (\autoref{sec:benchmark}), we find that smaller models without reasoning capability, such as Qwen3-8B and Qwen3-0.6B, do not always follow the prompts we provide. Consequently, they sometimes create actions that cannot be used in our Blender rendering pipeline.

\section{Survey Screenshots}
\begin{figure*}
    \centering
    \includegraphics[width=\linewidth]{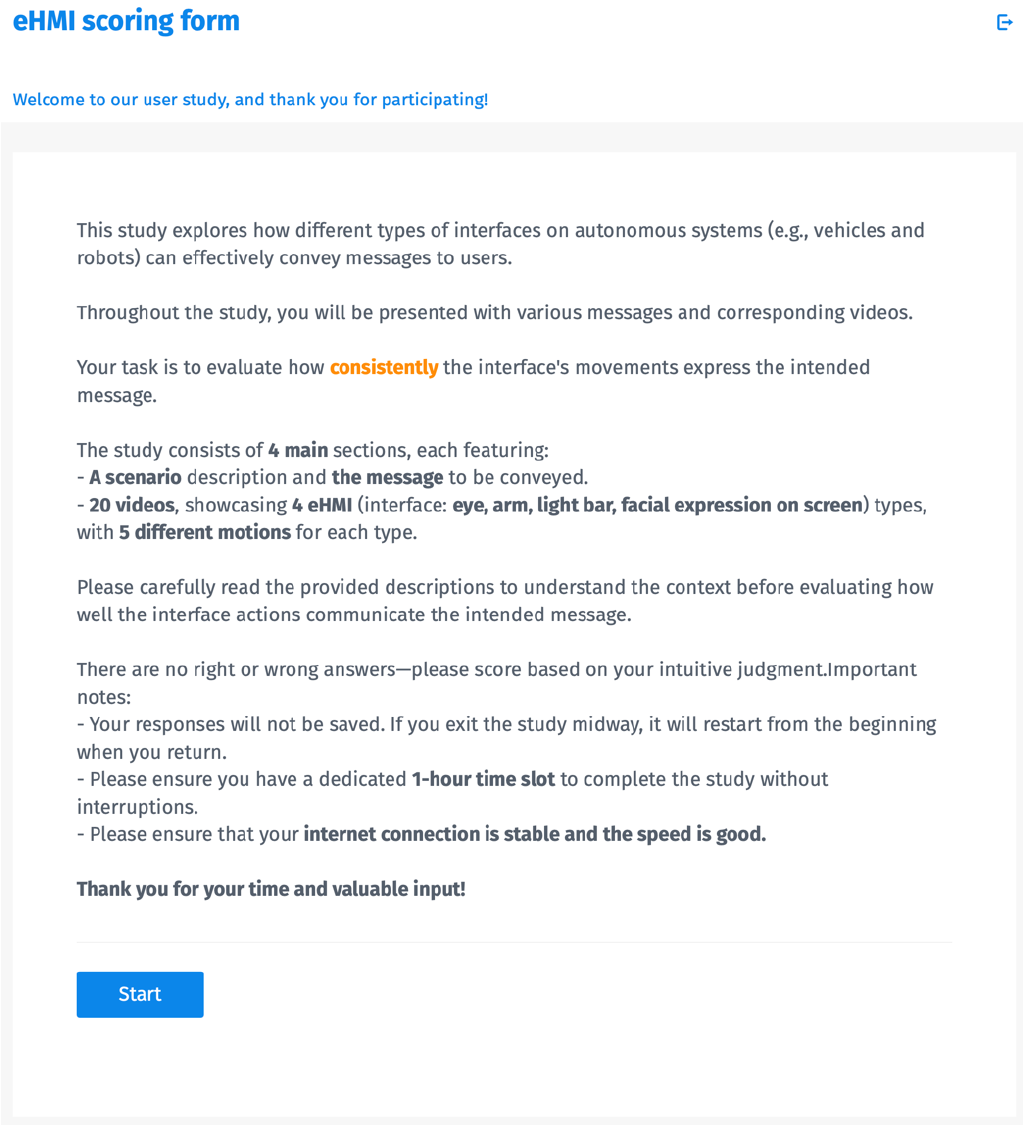}
    \caption{Introduction page of our action scoring survey}
    \label{fig:survey_intro_page}
\end{figure*}

\begin{figure*}
    \centering
    \includegraphics[width=\linewidth]{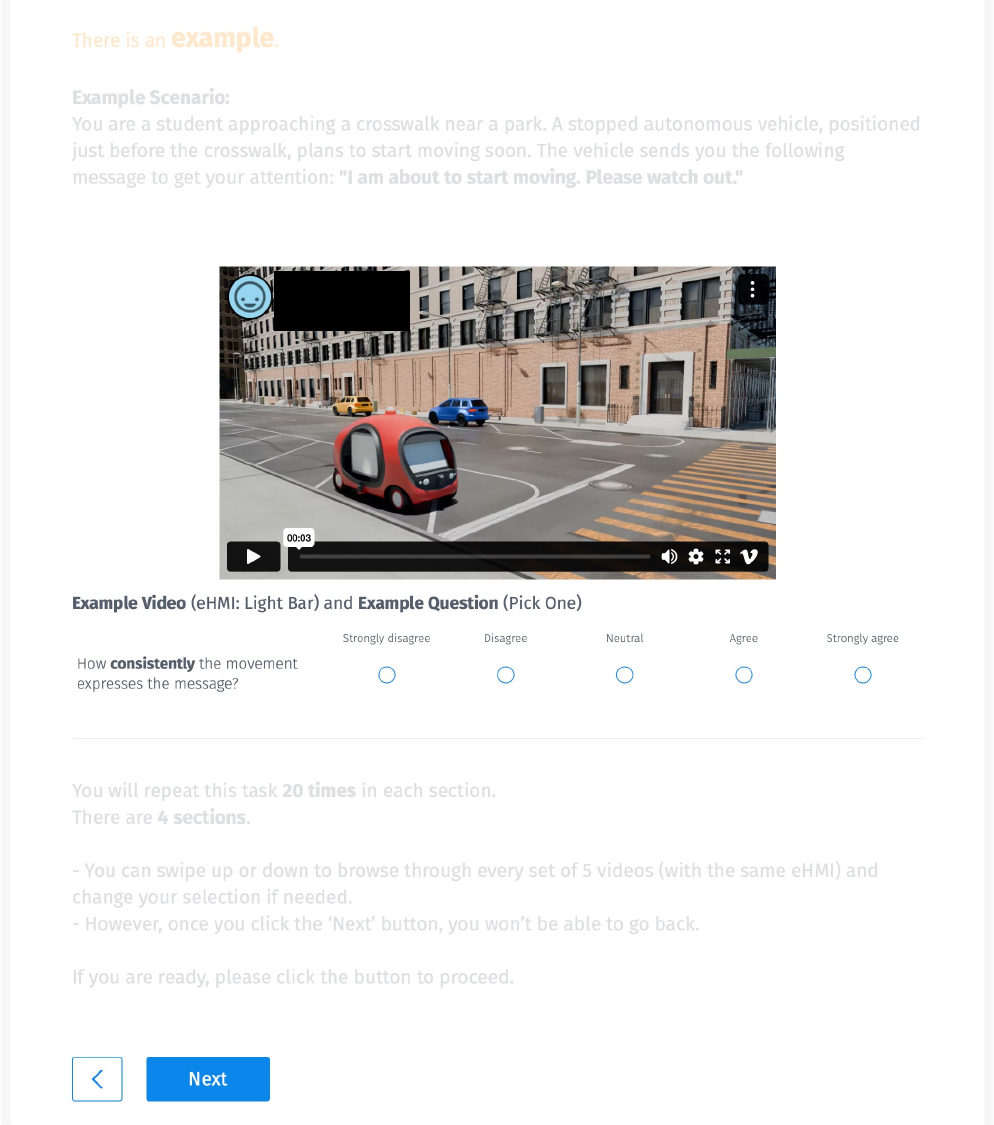}
    \caption{Demo page of our action scoring survey}
    \label{fig:survey_demo_page}
\end{figure*}

\begin{figure*}
    \centering
    \includegraphics[width=\linewidth]{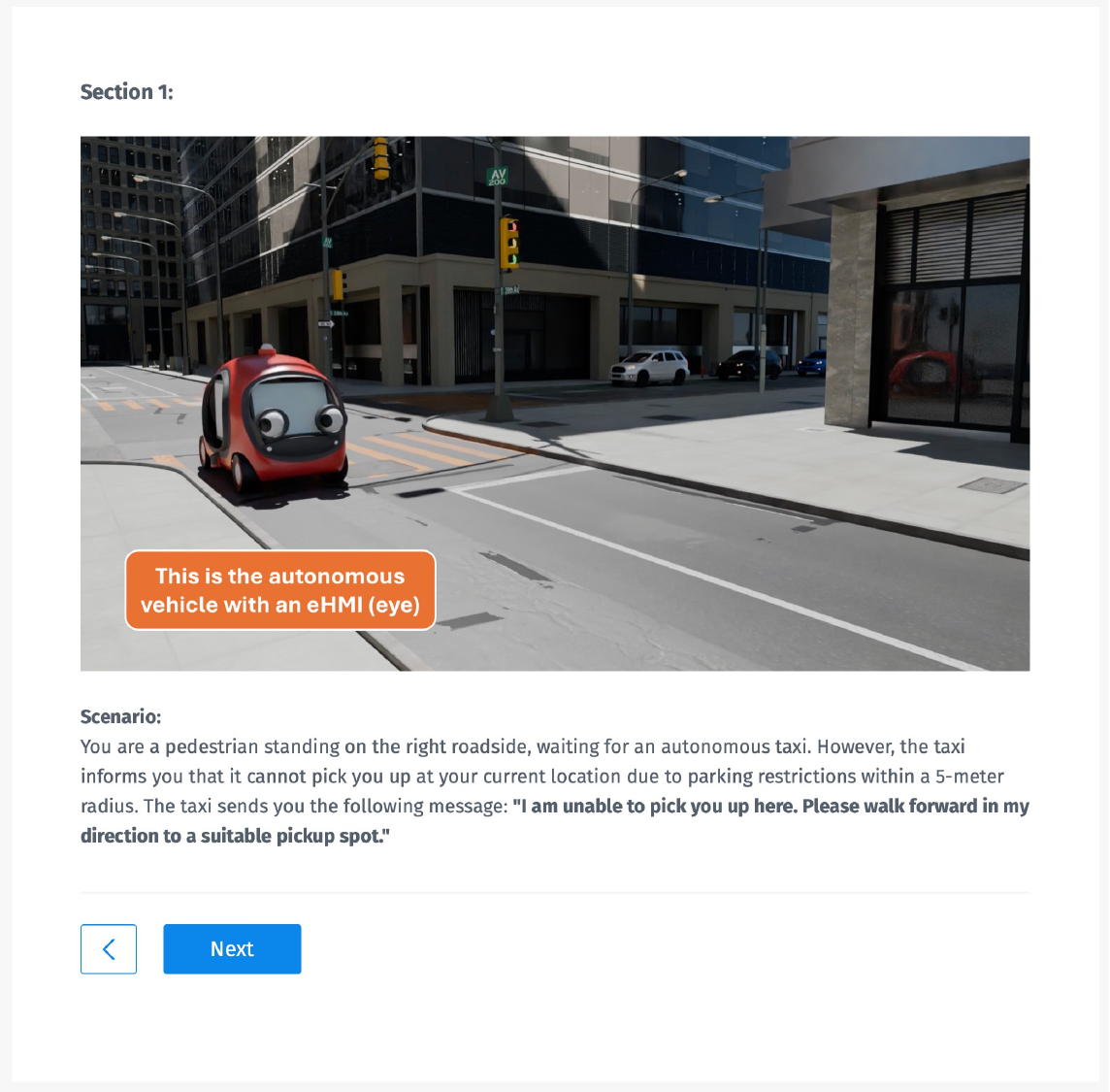}
    \caption{Scenario introduction page of our action scoring survey}
    \label{fig:survey_scenario_page}
\end{figure*}

\begin{figure*}
    \centering
    \includegraphics[width=\linewidth]{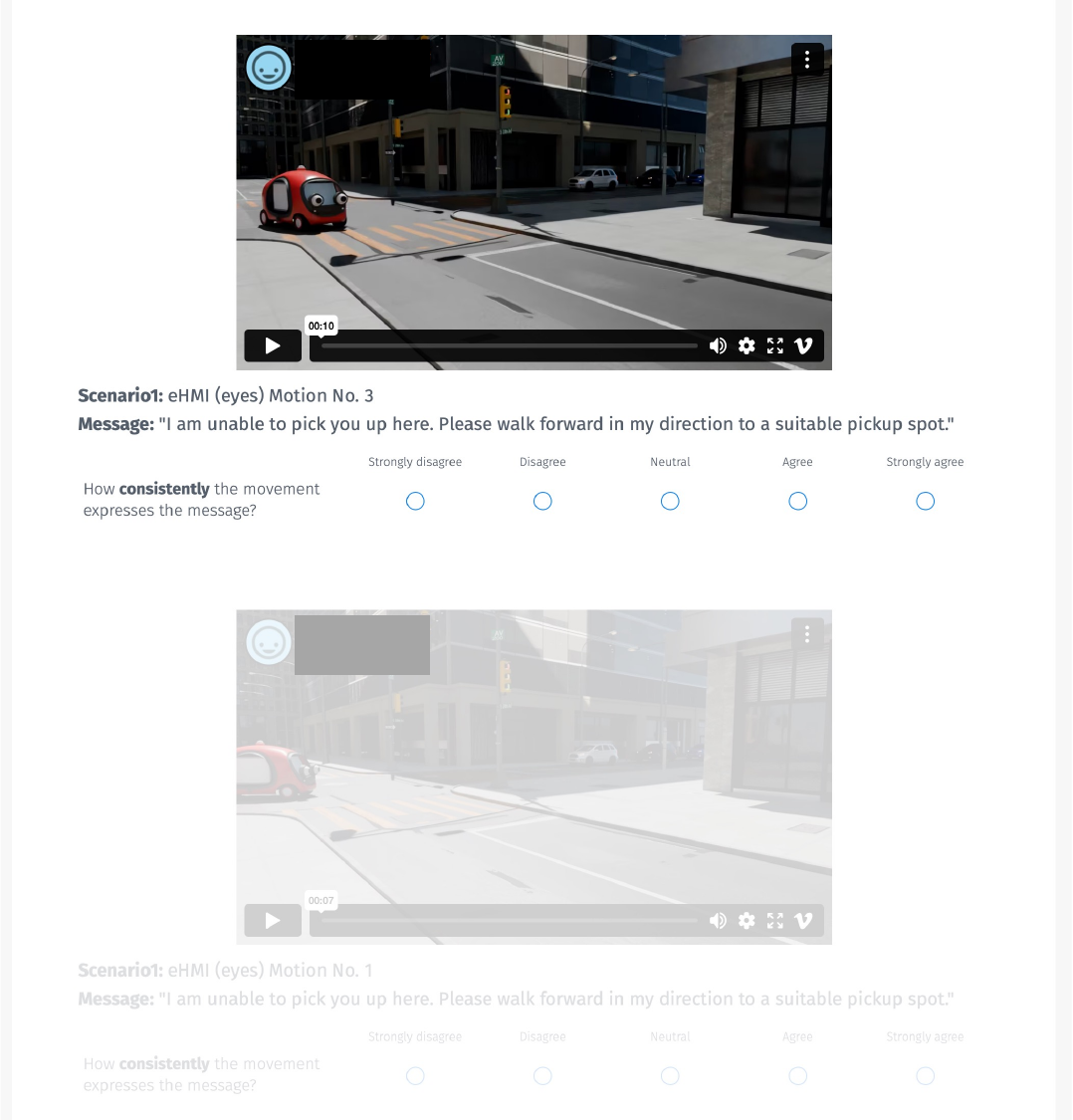}
    \caption{Participant rating page of our action scoring survey}
    \label{fig:survey_rating_page}
\end{figure*}

We provide detailed guidance for our data collection process. \autoref{fig:survey_intro_page} shows the introduction page of our survey. \autoref{fig:survey_demo_page} is a demonstration; \autoref{fig:survey_scenario_page} introduces the next rating scenario, and \autoref{fig:survey_rating_page} is the page participants use to rate clips.

% \newcommand{\widthscene}{0.47}
% \begin{figure*}
%     \small
%     \centering
%     \setlength{\tabcolsep}{3pt}
%     \begin{tabular}{cc}
%          \includegraphics[width=\widthscene\linewidth]{figures/scenario_demo_images/scenario_0_demo_image.pdf} &  \includegraphics[width=\widthscene\linewidth]{figures/scenario_demo_images/scenario_1_demo_image.pdf} \\
%          Scenario 0: Send intention &  Scenario 1: Status report  \vspace{0.5em} \\
%          \includegraphics[width=\widthscene\linewidth]{figures/scenario_demo_images/scenario_2_demo_image.pdf} &  \includegraphics[width=\widthscene\linewidth]{figures/scenario_demo_images/scenario_3_demo_image.pdf} \\
%          Scenario 2: Request help &  Scenario 3: Refuse help \vspace{0.5em} \\
%          \includegraphics[width=\widthscene\linewidth]{figures/scenario_demo_images/scenario_4_demo_image.pdf} &  \includegraphics[width=\widthscene\linewidth]{figures/scenario_demo_images/scenario_5_demo_image.pdf} \\
%          Scenario 4: Pedestrian Blind Spot Alert &  Scenario 5: Driver Blind Spot Warning \vspace{0.5em} \\
%          \includegraphics[width=\widthscene\linewidth]{figures/scenario_demo_images/scenario_6_demo_image.pdf} &  \includegraphics[width=\widthscene\linewidth]{figures/scenario_demo_images/scenario_7_demo_image.pdf} \\
%          Scenario 6: Target Identification &  Scenario 7: Broadcast Communication \\
%     \end{tabular}
%     \caption{Visualization of eight scenario-modality pairs we designed.}
%     \label{fig:vis_scene}
% \end{figure*}

\end{document}